\documentclass[twocolumn,prb,superscriptaddress]{revtex4}

\usepackage{graphicx}
\usepackage{dcolumn}
\usepackage{bm}
\usepackage{amsmath}
\usepackage{amsfonts}
\usepackage{array}
\usepackage{psfrag}
\newcommand{\vect}[1]{\ensuremath{\mathbf{ #1 }}}
\newcommand{\ket}[1]{\ensuremath{\left| #1 \right\rangle}}
\newcommand{\bra}[1]{\ensuremath{\left\langle #1 \right|}}
\newcommand{\gsbra}{\ensuremath{\bra{F}}}
\newcommand{\gsket}{\ensuremath{\ket{F}}}
\newcommand{\braket}[2]{\ensuremath{\left\langle #1 \, \right| \left. \! #2 \right\rangle}}

\newcommand{\commute}[2]{\ensuremath{\left[ #1 , #2 \right]}}

\newcommand{\pauli}[5]{\ensuremath{ \lambda_{#3 , #4}^{#1 , #2} (\mathbf{#5}) }}
\newcommand{\direct}[4]{\ensuremath{ \xi_{#3 , #4}^{#1 , #2} }}
\newcommand{\Nel}{\ensuremath{ N_{\text{el}} }}
\newcommand{\nel}{\ensuremath{ n_{\text{el}} }}
\newcommand{\Nx}{\ensuremath{ N_{\text{exc}} }}

\newcommand{\nqw}{\ensuremath{ n_{\text{QW}} }}

\newcommand{\sumk}[1]{\sum_{\mathbf{#1}}}
\newcommand{\Eq}[1]{Eq. (\ref{#1})}
\newcommand{\ca}[2]{c_{#1,\mathbf{#2}}}
\newcommand{\cc}[2]{c^{\dagger}_{#1, \mathbf{#2}}}
\newcommand{\pa}[1]{a_{\mathbf{#1}}}
\newcommand{\pc}[1]{a^{\dagger}_{\mathbf{#1}}}
\newcommand{\ba}[2]{b_{#1,\mathbf{#2}}}
\newcommand{\bc}[2]{b^{\dagger}_{#1,\mathbf{#2}}}
\newcommand{\Ba}[1]{B_{\mathbf{#1}}}
\newcommand{\Bc}[1]{B^{\dagger}_{\mathbf{#1}}}
\newcommand{\Ta}[1]{T_{\mathbf{#1}}}
\newcommand{\Tc}[1]{T^{\dagger}_{\mathbf{#1}}}
\newcommand{\ta}[1]{t_{\mathbf{#1}}}
\newcommand{\tc}[1]{t^{\dagger}_{\mathbf{#1}}}
\newcommand{\nuc}[2]{\nu_{#1,\mathbf{#2}}}

\newcommand{\upa}[1]{p_{\text{U} \, \mathbf{#1}}}
\newcommand{\lpc}[1]{p^{\dagger}_{\text{L} \, \mathbf{#1}}}
\newcommand{\lpa}[1]{p_{\text{L} \, \mathbf{#1}}}
\newcommand{\alphaHopf}[1]{\alpha_{\mathbf{#1}}}
\newcommand{\betaHopf}[1]{\beta_{\mathbf{#1}}}

\begin{document}
\title{Many-body physics of intersubband polaritons}

\author{Luc \surname{Nguyen-th\^e}}
\affiliation{Laboratoire Mat\'eriaux et Ph\'enom\`enes Quantiques, Universit\'e Paris  Diderot-Paris 7 and CNRS, UMR 7162, 75013 Paris, France}
\author{Simone \surname{De Liberato}}
\affiliation{Laboratoire Mat\'eriaux et Ph\'enom\`enes Quantiques, Universit\'e Paris  Diderot-Paris 7 and CNRS, UMR 7162, 75013 Paris, France}
\affiliation{School of Physics and Astronomy, University of Southampton, Southampton, SO17 1BJ, United Kingdom}
\author{Motoaki Bamba}
\affiliation{Laboratoire Mat\'eriaux et Ph\'enom\`enes Quantiques, Universit\'e Paris  Diderot-Paris 7 and CNRS, UMR 7162, 75013 Paris, France}
\affiliation{Department of Physics, Osaka University, 1-1 Machikaneyama, Toyonaka, Osaka 560-0043, Japan}
\author{Cristiano Ciuti}
\affiliation{Laboratoire Mat\'eriaux et Ph\'enom\`enes Quantiques, Universit\'e Paris  Diderot-Paris 7 and CNRS, UMR 7162, 75013 Paris, France}
 
\begin{abstract}
Intersubband polaritons are light-matter excitations originating from the strong coupling between an intersubband quantum well electronic transition and a microcavity photon mode.
In this paper we study how the Coulomb electron-electron interaction and the Pauli saturation of the electronic transitions affect the physics of intersubband polaritons.
We develop a microscopic many-body theory for the physics of such composite bosonic excitations in a microcavity-embedded two-dimensional electron gas. As a first application, we calculate the modification of the depolarization shifts and the efficiency of intersubband polariton-polariton scattering processes.
\end{abstract}

\maketitle
\section{Introduction}

Intersubband polaritons are elementary excitations of microcavities embedding doped semiconductor quantum wells. In particular, they are the normal modes due to the strong coupling between a cavity photon mode and
the transition between two conduction subbands in quantum wells containing a two-dimensional electron gas. 
Intersubband polaritons were experimentally demonstrated for the first time in 2003\cite{Dini03} and their physics is the object of many interesting investigations, both fundamental
\cite{Gunter09,Ciuti05,DeLiberato07,Dupont07,Pereira07,Sapienza08,Todorov09,Anappara09,Zaluzny09,Jouy10,Auer12} and applied \cite{Colombelli05,Anappara05,Anappara06,Plumridge08,Todorov08,Geiser10,DeLiberato12b,Delteil11,Porer12}. 

So far, theoretical investigations have been based mostly on a bosonized Hamiltonian approach \cite{Ciuti05}. In the dilute regime, i.e., when the excitation density is much smaller than the electronic density in the doped quantum wells, intersubband polaritons behave as (composite) bosons. At higher excitation densities anyway, Pauli saturation effects due to electron fermionic statistics and deviations from the bosonic behavior are expected \cite{DeLiberato09}. 

An important aspect of the physics of such excitations is the role of the Coulomb interaction. In the case of bare intersubband transitions,  Coulomb interaction is known not to modify dramatically the polaritonic lineshape \cite{Nikonov97}, being responsible mainly for a simple shift of the excitation frequency, the so-called depolarization shift.
This is why, in many relevant situations, the presence of the Coulomb interaction can be in first approximation neglected, by using a renormalized (measured) value for the frequency of the intersubband transition instead of the bare one, calculated from the single-electron subband structure. 

Very recently,  the role of Coulomb interaction in intersubband polariton systems has started to attract interest. 
In Ref. [\onlinecite{DeLiberato12}] the part of the Coulomb interaction responsible for the depolarization shift has been studied in the bosonic excitation formalism, allowing to calculate the renormalization from first principles.
Other works have studied the same problem using different gauges \cite{Todorov10,Todorov12} and a Green functions approach \cite{Shelykh12}.
As a matter of fact, up to now, only depolarization or static effects have been taken into account.
Hence, the general problem of the many-body physics of intersubband polaritons is largely unexplored, in particular the physics of polariton-polariton interactions and polariton-polariton scattering. 

In this paper we derive a microscopic many-body theory describing the intersubband polariton-polariton scattering due both to the Coulomb interaction and to the non-perfect bosonicity of the intersubband excitations. 
As a first application of the present theoretical formalism, we calculate how the depolarization shifts are modified and, most importantly, we determine the efficiency of the intersubband polariton-polariton scattering processes. 

This article is organized as follows. In Sec. \ref{sec:general}, we present the physical system, introduce the fermionic electron-photon Hamiltonian including the different channels of the (screened) Coulomb interaction. Moreover, we write the intersubband excitations in second quantized formalism. In Sec. \ref{coboson}, the composite boson commutator approach \cite{Combescot08} to calculate many-body matrix elements is presented for the case of intersubband excitations.  In Sec. \ref{effective}, an effective bosonic Hamiltonian (up to the fourth order in the bosonic operators) is derived in such a way that the two-excitation properties are the same as those predicted by the microscopic composite boson approach. In Sec.~\ref{numerical} numerical applications of the present theory to the calculation of the interaction energy between intersubband excitations are presented. Sec. \ref{scattering} is devoted to the calculation of polariton-polariton scattering processes. Finally, conclusions are drawn in Sec. \ref{conclusions}. Additional technical details
are included in Appendix \ref{sec:notation}, \ref{sec:coeff} and \ref{sec:fermi}.

 %%%%%%%%%%%%%%%
\section{The system and its Hamiltonian description}
\label{sec:general}

%%%%%%
\subsection{Physical system}

\begin{figure}[t!]
\begin{center}
\includegraphics[width=8.8cm]{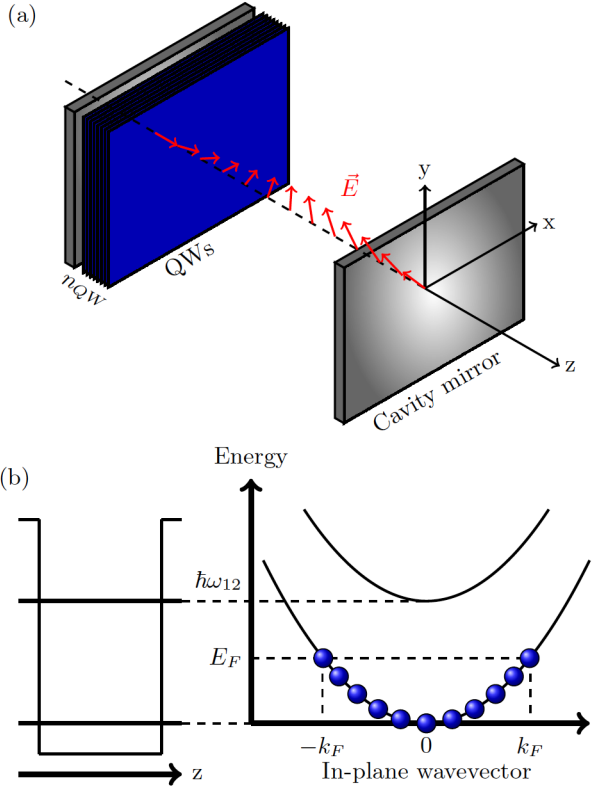}
\caption{\label{Sketch} Schematic representation of the system under investigation.
Panel (a): Multiple doped quantum wells (QWs) are embedded in a planar photonic microcavity. The image above shows also the fundamental TM photonic mode in the case of metallic mirrors. Due to intersubband selection rule, the intersubband transitions only couple to the $z$-component of the electric field.
Panel (b): Energy profile of the quantum well along the $z$-direction and its two bound states (left). Electronic dispersion of the first two conduction subbands in each of the doped quantum wells, as a function of the in-plane electronic wavevector (right). The Fermi wavevector, $k_F$, the Fermi Energy, $E_F$, and the intersubband gap, $\hbar\omega_{12}$, are highlighted.}
\label{fig:cavity}
\end{center}
\end{figure}

We will consider a photonic, planar microcavity embedding $\nqw$ symmetric quantum wells, each doped with an areal electron density of $\nel$ (see Fig. \ref{Sketch} for a schematic representation of such a system).
As the quantum wells confine the electrons along the growth direction, the electronic bands will be split into multiple, almost parallel subbands. The optically active transitions we are interested in will thus be the ones from the last occupied conduction subband to the first unoccupied one.
The photonic microcavity width will be chosen to have one of the photonic modes almost resonant with the electronic transition between these two electronic conduction subbands (the first two for simplicity, even if  more general situations are possible \cite{Anappara07}). In order to simplify the notation, we will consider that the quantum wells are all identically coupled to the relevant microcavity field.

%%%%%
% figure
%%%%%%

The Hamiltonian describing the coupled electron-photon system can be written in the Coulomb gauge as
\begin{eqnarray}
\label{H}
H&=&H_{C}+H_{F},
\end{eqnarray}
where $H_{C}$ includes only photonic degrees of freedom, while 
$H_F$ regroups all the terms presenting also electronic ones. 
Introducing the annihilation operators $\ca{\mu}{k}$ for an electron in the subband $\mu$ with an in-plane wavevector $\mathbf{k}$, and $\pa{q}$ for a microcavity photon with an in-plane wavevector $\mathbf{q}$, we can write the two Hamiltonians explicitly as 
\begin{eqnarray}
\label{HF}
	H_{C} & = & \sumk{q} \hbar\omega_{c,q}\pc{q}\pa{q}+ \hbar \Delta_q (\pa{q}+\pc{-q}) (\pa{-q}+\pc{q})\nonumber \\
	H_{F} & = & H_{\text{Kin}} + H_{\text{Rabi}} + H_{\text{Coul}},
	\label{eq:defH}
\end{eqnarray}
where
\begin{eqnarray}
\label{eq:HFdetail}
	H_{\text{Kin}} & = & \sum_{\mu=\{1,2\},\mathbf{k}} \hbar\omega_{\mu,k}\cc{\mu}{k}\ca{\mu}{k}  \\
	H_{\text{Rabi}} & = & \sumk{k,q} \hbar\chi_q (\cc{2}{k+q}\ca{1}{k}+\cc{1}{k+q}\ca{2}{k})(\pa{q}+\pc{-q}) \nonumber \\
	H_{\text{Coul}} & = & \frac{1}{2} \sum_{ \substack{ \mathbf{k, k' ,q} \\ \mu, \mu' , \nu, \nu' \in \{ 1,2\}} }  V_{ q }^{ \mu \nu \nu' \mu' }  \cc{\mu}{k+q} \cc{\nu}{k'-q} \ca{\nu'}{k'} \ca{\mu'}{k},\nonumber
\end{eqnarray}
are respectively the electronic kinetic energy, the light-matter coupling Hamiltonian and the electron-electron Coulomb interaction.
Coefficients $\omega_{c,q}$ and $\omega_{\mu,k}$ are the frequencies of a photon with in-plane wavevector $\mathbf{q}$ and of an electron with in-plane wavevector $\mathbf{k}$ in the subband $\mu$, $\chi_q$ is the light-matter coupling frequency, $\Delta_q$ the diamagnetic interaction coefficient and the $\mathbf{q}=0$ term has to be excluded from the sum in  $H_{\text{Coul}} $ in order to compensate for the homogeneous positive background of the jellium model \cite{Giuliani05}.  

In order to simplify the notation both spin and quantum well indexes have been omitted but one should keep in mind that there is an implicit sum over them. The explicit notation is given in Appendix~\ref{sec:notation}.
We also neglect to explicitly write the photon polarization because, due to the intersubband selection rule, only the TM modes are coupled with matter.

The Coulomb matrix elements $V^{\mu\nu\nu'\mu'}_q$  are given by \cite{Lee99}
\begin{eqnarray}
\label{V}
	V^{\mu\nu\nu'\mu'}_q&=&\frac{e^2}{2 S \epsilon_0\epsilon_{r} q} I_{q}^{\mu\nu\nu'\mu'} \\
	I_{q}^{\mu\nu\nu'\mu'} & = & \int dz dz' \psi_{\mu}(z)\psi_{\nu}(z)  \psi_{\nu'}(z')\psi_{\mu'}(z') e^{-q\lvert z-z' \lvert} ,\nonumber
\end{eqnarray}
where $S$ is the surface of the sample and $\psi_{\mu}(z)$ is the real $z$-component  wavefunction of an electron in the subband $\mu$.

\begin{figure*}[t!]
\begin{center}
\begin{tabular}{rrrr}
	\includegraphics[width=4cm]{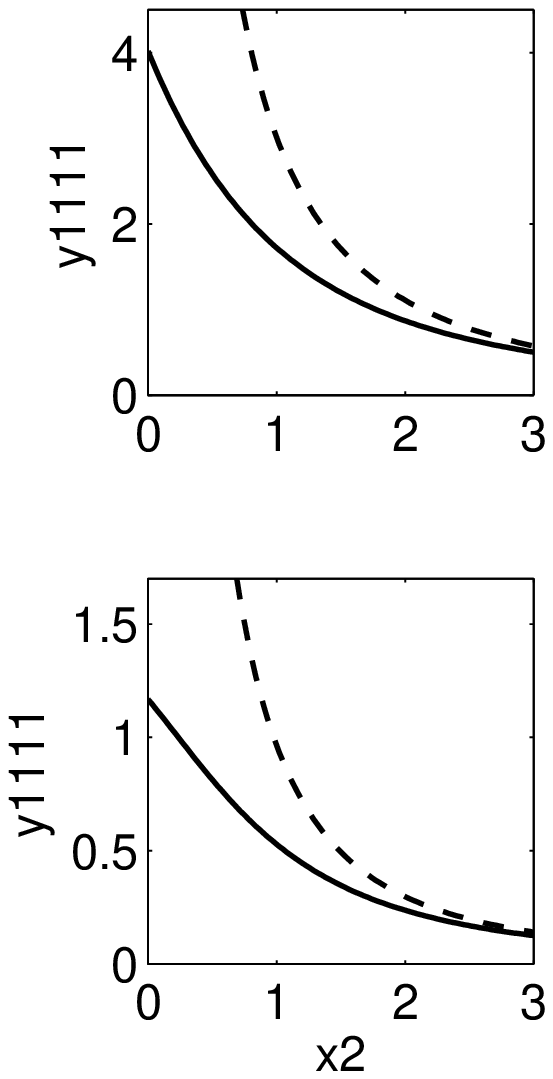} & \includegraphics[width=4cm]{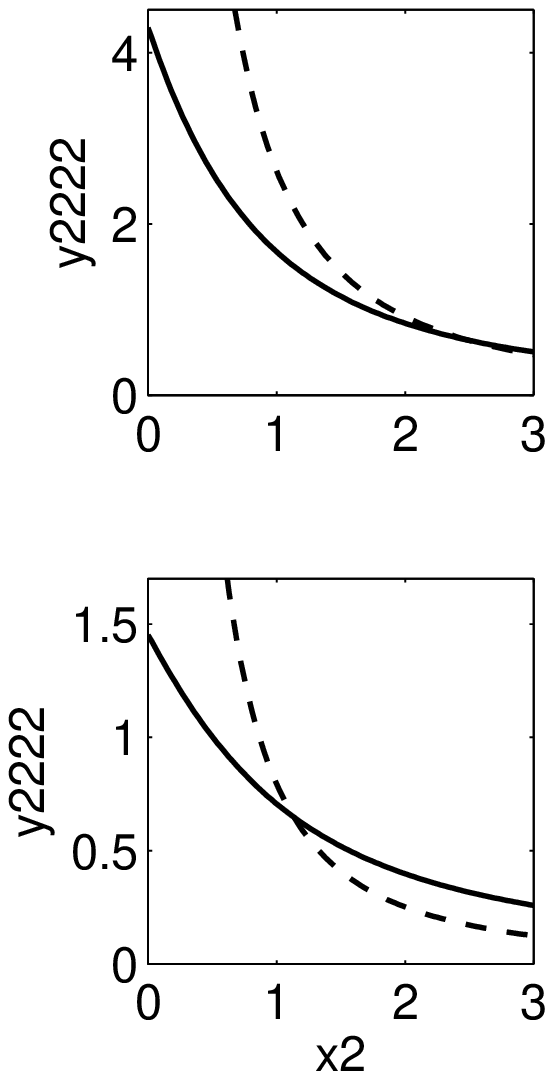} & \includegraphics[width=4cm]{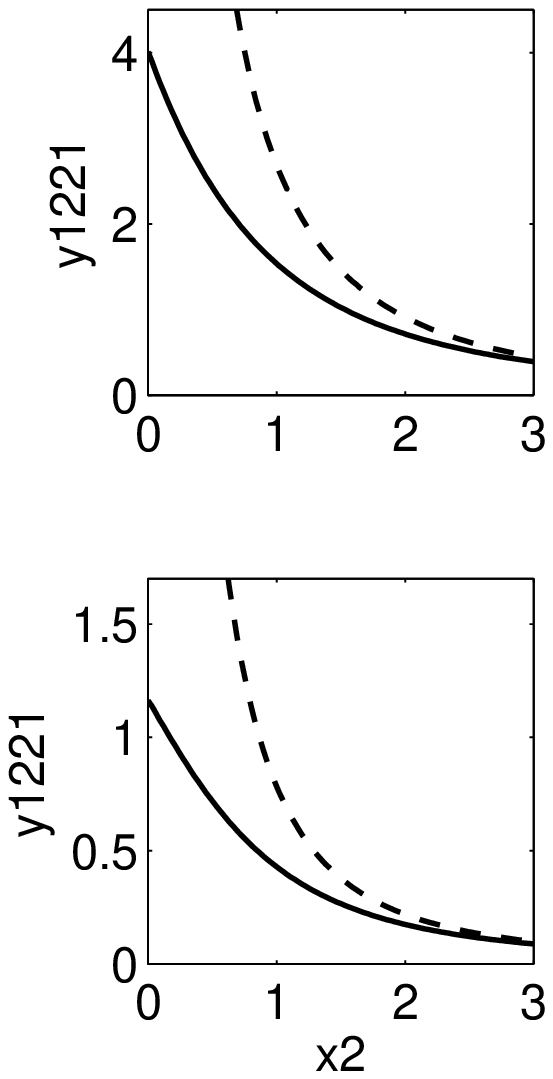} & \includegraphics[width=4cm]{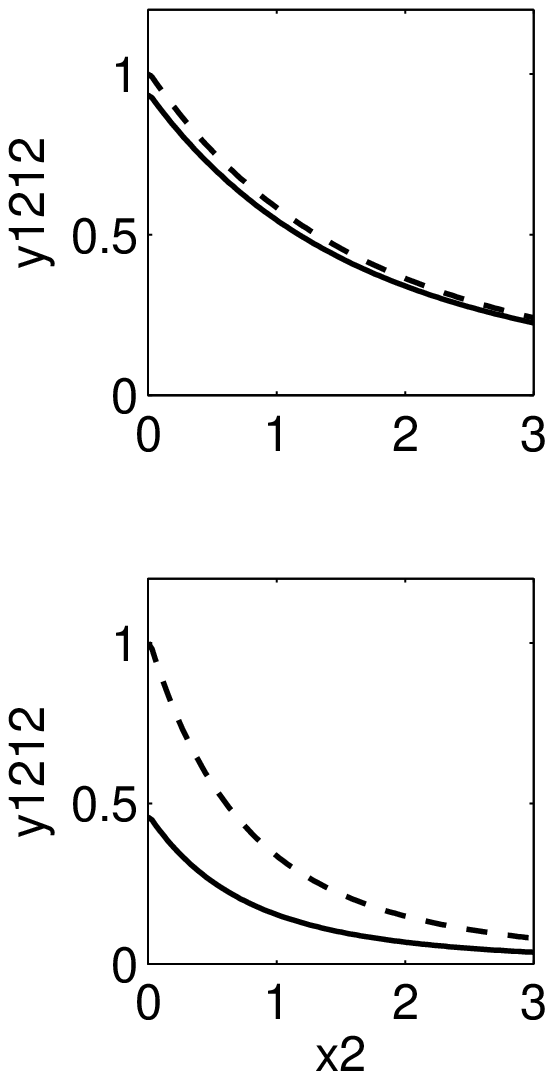} \\
	\includegraphics[width=4cm]{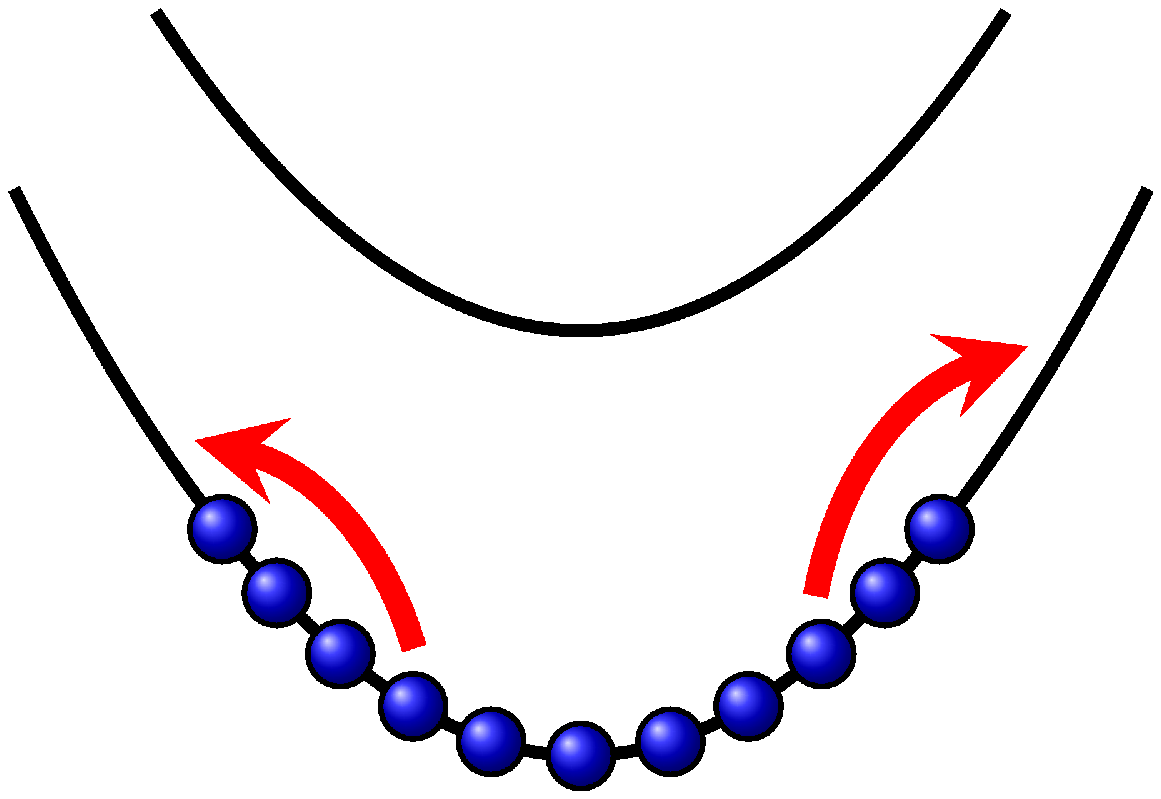} & \includegraphics[width=4cm]{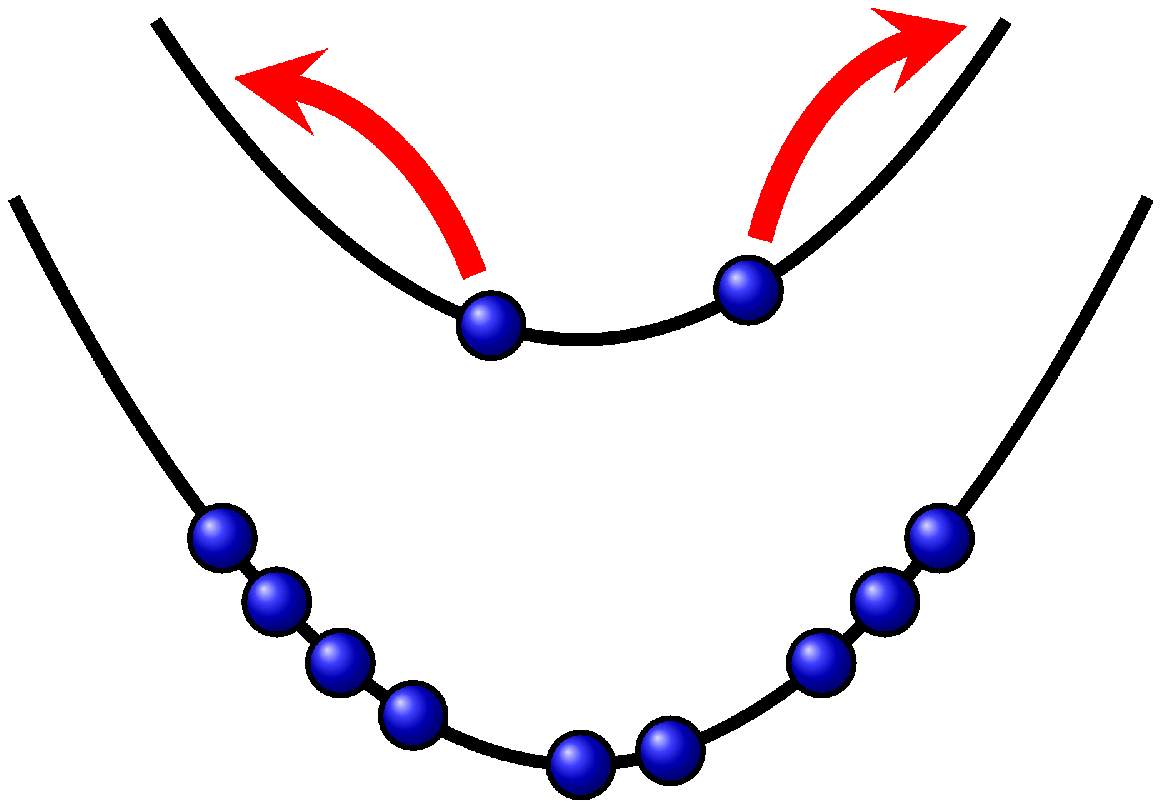} & \includegraphics[width=4cm]{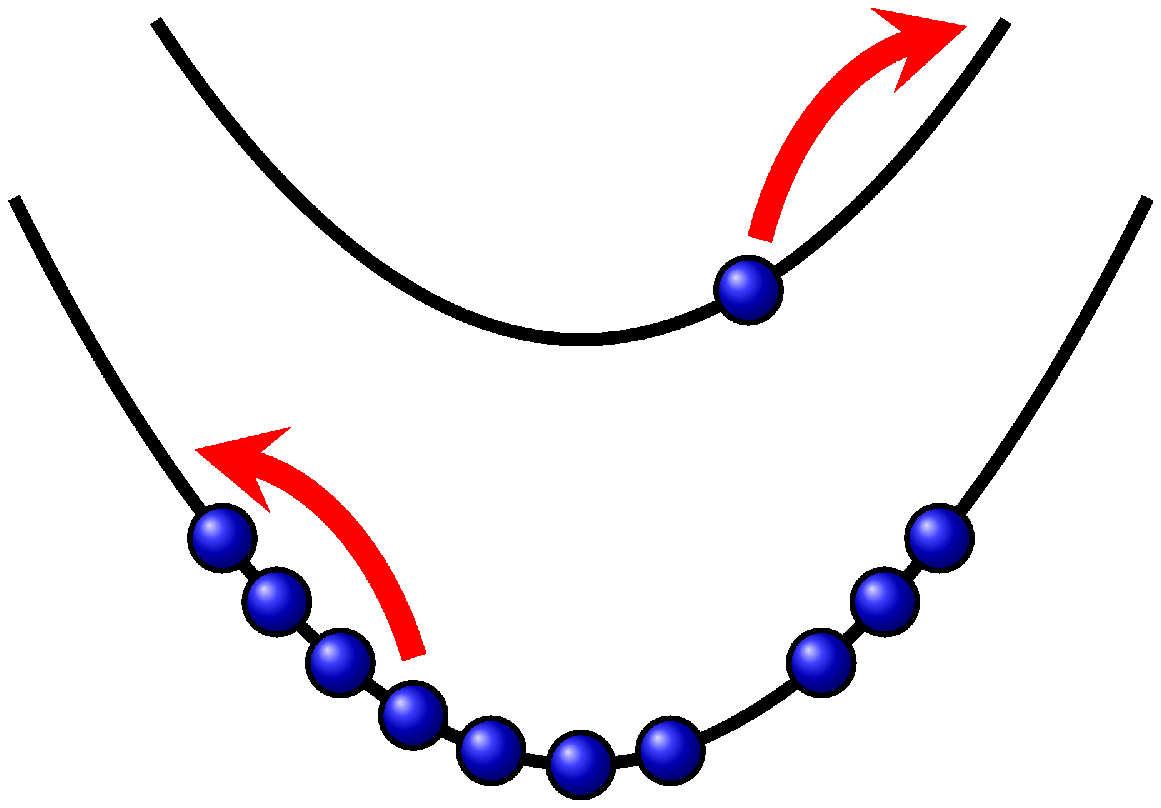} & \includegraphics[width=4cm]{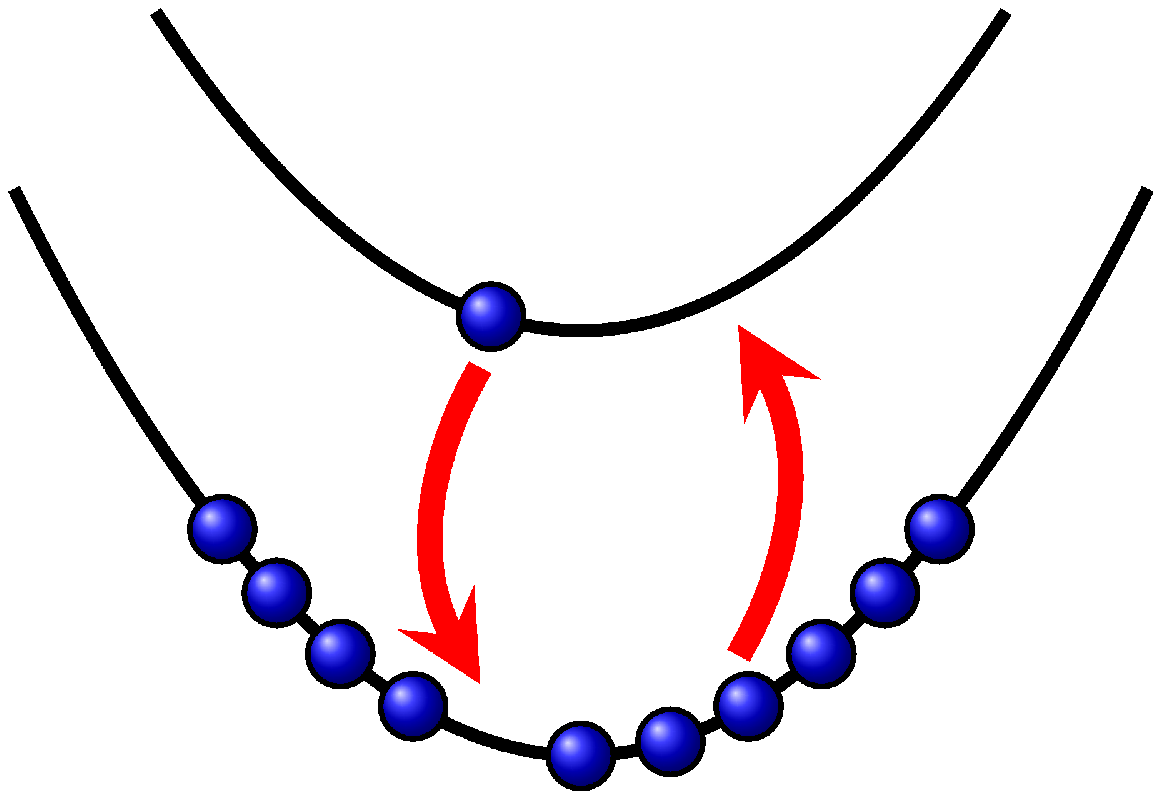}
\end{tabular}
\caption{ Panels in the first two lines show the wavevector dependency of the 
the four qualitatively different Coulomb processes. In each panel we plot both the bare potentials (dashed line) and the static-RPA screened ones (solid line). The first row is for a mid-infrared transition, with $\hbar\omega_{12}=140$meV  and an electronic density in each quantum well $\nel = 5 \times 10^{11} \text{ cm}^{-2}$. The second row is instead for a THz transition, with $\hbar\omega_{12}=15$meV and $\nel = 2 \times 10^{11} \text{ cm}^{-2}$. 
In the third row we present schematic representation of the four relevant processes. }
\label{fig:coulomb}
\end{center}
\end{figure*}

Due to the symmetry of the quantum wells, and thus of the wavefunctions, a certain number of matrix elements in \Eq{V} can be seen to be zero, in particular all matrix elements with an odd number of $1$ and $2$ indices
\begin{eqnarray}
\label{elem0}
V^{1112}_q&=& V^{1121}_q= V^{1211}_q= V^{2111}_q=0\\
V^{2111}_q&=& V^{2212}_q= V^{2122}_q= V^{1222}_q=0.\nonumber 
\end{eqnarray}
The other elements obey the relations
\begin{eqnarray}
\label{elem}
V^{1122}_q&=& V^{1212}_q= V^{2121}_q= V^{2211}_q\\
V^{1221}_q&=& V^{2112}_q\nonumber,
\end{eqnarray}
so that there are only four  distinct matrix elements, $V^{1212}_q$,$V^{1221}_q$,$V^{1111}_q$ and $V^{2222}_q$. 
In Fig.~\ref{fig:coulomb} we show a graphical representation of these four qualitatively different processes and the wavevector dependency of their respective coefficients
for an infinite potential well. From such a figure we can clearly see a qualitative difference  in the long wavelength behavior between  $V^{1212}_q$ and the other three coefficients.
As explained in greater detail in Ref. [\onlinecite{DeLiberato12}],  they indeed refer  to qualitatively different physical processes.
While $V^{1212}_q$ describes an intersubband, coherent Coulomb scattering, that is responsible for the so-called depolarization shift, the others describe intrasubband scattering processes, in which electrons are scattered inside the same subband.

%%%%%
% figure
\psfrag{x2}[bc]{$q/k_{F}$}
\psfrag{y1111}[Bc]{$\quad\quad V_{q}^{1111}$ (a.u.)}
\psfrag{y2222}[Bc]{$\quad\quad V_{q}^{2222}$ (a.u.)}
\psfrag{y1221}[Bc]{$\quad\quad V_{q}^{1221}$ (a.u.)}
\psfrag{y1212}[Bc]{$\quad\quad V_{q}^{1212}$ (a.u.)}
%%%%%%%

%%%%%%
\subsection{Screened Coulomb interaction}

The dense two dimensional electron gas in the first electronic subband screens the Coulomb interaction. In order to take it into account we will thus replace the bare Coulomb interaction  ${V}^{\mu\nu\nu'\mu'}_{q}$  defined by \Eq{V}, with its static RPA-screened version
 $\tilde{V}^{\mu\nu\nu'\mu'}_{q}$. 
From Refs. [\onlinecite{Lee99,Lee00}], these coefficients obey the Dyson equation
\begin{eqnarray}
	\tilde{V}^{\mu\nu\nu'\mu'}_{q} & = &
	V^{\mu\nu\nu'\mu'}_{q}  \\
	&& + \sum_{\alpha,\beta}
	V^{\mu\beta\alpha\mu'}_{q} \Pi^{\alpha\beta}(\mathbf{q},\omega=0) \tilde{V}^{\alpha\nu\nu'\beta}_{q} ,   \nonumber
\end{eqnarray}
where $\Pi^{\alpha\beta}(\mathrm{q},\omega)$ is the RPA polarization function for the $\alpha\rightarrow\beta$ transition.
The screened interactions thus take the form
\begin{eqnarray}
	\tilde{V}^{1\nu\nu1}_{q} & = &
	\frac{V^{1\nu\nu1}_{q}}{1-V^{1111}_{q}\Pi^{11}(\mathbf{q},0)}   \\
	\tilde{V}^{2222}_{q} & = &
	\frac{V^{2222}_{q} + \lbrack (V^{1221}_{q})^{2} - V^{1111}_{q}V^{2222}_{q}\rbrack\Pi^{11}(\mathbf{q},0)}{1-V^{1111}_{q}\Pi^{11}(\mathbf{q},0)} \nonumber \\
	\tilde{V}^{1212}_{q} & = &
	\frac{V^{1212}_{q}}{1-V^{1212}_{q}[ \Pi^{12}(\mathbf{q},0)+\Pi^{21}(\mathbf{q},0) ]},\nonumber
	 \label{eq:screen22}
\end{eqnarray}
where $\nu = \{1,2\}$.
Screening of intrasubband and intersubband processes can thus be encoded respectively in the dielectric functions $\epsilon(\mathbf{q})$ and $\epsilon_{12}(\mathbf{q})$, where
\begin{eqnarray}
	\epsilon(\mathbf{q}) = 1-V^{1111}_{q}\Pi^{11}(\mathbf{q},0) ,
\end{eqnarray}
and
\begin{eqnarray}
	\epsilon_{12}(\mathbf{q}) = 1-V^{1212}_{q}[ \Pi^{12}(\mathbf{q},0)+\Pi^{21}(\mathbf{q},0) ] .
	\label{eq:dielecDepol}
\end{eqnarray}
As we will see in the rest of this Section, only the long-wavelength limit is relevant in the study of intersubband polaritons. We will thus always consider the $\mathbf{q}\rightarrow 0$ limit of the above experessions.
In Fig. \ref{fig:coulomb} the reader can find the screened and unscreened profiles of the different potentials as a function of the exchanged momentum normalized over the Fermi wavevevector $q/k_F$.

%%%%%%
\subsection{Intersubband excitations}
\label{sec:ISB}

The theory in this paper will be developed using a zero temperature ($T=0$) formalism. In order for this approximation to be valid, we need to have a vanishing thermal electronic population in the excited subband, that is $\hbar{\omega}_{2,0}-E_F\gg k_BT$, where $E_F$ is the Fermi energy and $k_B$ the Boltzmann constant.
Under such a condition the ground state of the uncoupled light-matter system is given by
\begin{eqnarray}
\ket{F} &=& \prod_{|\vect{k}| \leq k_{F}} \cc{1}{k} \ket{0},
\end{eqnarray}
where $k_F$ is the Fermi wavevector and $\ket{0}$ is the vacuum state, with no electrons in both subbands and no photons in the microcavity. In the ground state all electronic states up to the Fermi energy are occupied and all the others are empty.

If we neglect the incoherent intrasubband Coulomb processes (due to the coefficients $V_{q}^{2222}$, $V_{q}^{1221}$ and $V_{q}^{1111}$) in \Eq{eq:HFdetail}, the full Hamiltonian $H$ can be solved, in the dilute regime, in terms  of bosonic coherent excitations named intersubband polaritons\cite{Ciuti05}.

In order to do this, we first notice that the microcavity photon mode is coupled through $H_{\text{Rabi}}$ to only one particular linear superposition of electronic transitions
\begin{eqnarray}
\label{eq:isb0}
	\bc{0}{q} & = &
	\frac{1}{ \sqrt{\nqw \Nel} }
	\sumk{k} \nuc{0}{k}^{\ast} \, \cc{2}{k+q} \ca{1}{k}  ,
\end{eqnarray}
where $\nuc{0}{k} = \Theta(k_F - k)$ and $\Theta$ is the Heaviside function.
The choice of the notation with the index $0$ will be explained in the following. In \Eq{eq:isb0}, as well as in \Eq{eq:HFdetail}, the sum is implicit over the spin and quantum well indexes to simplify notations. This defines an intersubband collective excitation, which we will name bright intersubband excitation. Hamiltonian $H_{\text{Rabi}}$ can thus be written exactly
\begin{eqnarray}
\label{eq:Hrabi}
	H_{\text{Rabi}} & = &
	\sumk{q} \hbar\Omega_q (\ba{0}{-q}+\bc{0}{q})(\pa{q}+\pc{-q}) ,
\end{eqnarray}
where $\Omega_q=\sqrt{\nqw \Nel}\chi_q$ is the Rabi frequency. \Eq{eq:Hrabi} makes it clear that bright excitations concentrate all the oscillator strength of the electron gas.
Other similar collective excitations can be constructed by an orthonormalization procedure
\begin{eqnarray}
\label{isbt}
	\bc{i}{q} &=&
	\frac{1}{\sqrt{\nqw \Nel }} \sumk{k}  \nuc{i}{k}^{\ast} \, \cc{2}{k+q} \ca{1}{k},
\end{eqnarray}
where index $i$ runs from $1$ to $\nqw\Nel-1$ and the $\nu$ coefficients have support over the Fermi sea and satisfy the orthonormality relation
\begin{eqnarray}
\label{eq:orthonormal}
	\frac{1}{\nqw \Nel}  \sumk{k} \nuc{i}{k}^{\ast} \nuc{j}{k} & = &  \delta_{i,j}.
\end{eqnarray}
However, none of these new collective excitations is coupled to the microcavity photon field. This is why we call them dark intersubband excitations. We will consider them in Sec. \ref{coboson}.

We then use the fact that the two electronic subbands are parallels, 
and we neglect the photonic exchanged momentum when compared to an
electronic one, that is
\begin{eqnarray}
\label{parallel}
\omega_{2,k}&=&\omega_{1,k}+\omega_{12} \nonumber \\
\omega_{\mu,\lvert \mathbf{k+q}\lvert}&\simeq&\omega_{\mu,k}.
\end{eqnarray}
We can thus rewrite the Hamiltonian as \cite{Ciuti05,DeLiberato12}
\begin{eqnarray}
	\label{HIB}
	\tilde{H}&=&H_C+\sum_{j,\mathbf{k}}\hbar\omega_{12}\bc{j}{q}\ba{j}{q}\\&&
	+ \sumk{q} \hbar\Omega_q (\ba{0}{-q}+\bc{0}{q})(\pa{q}+\pc{-q})\nonumber \\&&
	+  \frac{\Nel}{2} \sumk{q} \tilde{V}_{q}^{1212} \left( \bc{0}{q} + \ba{0}{-q} \right) \left( \bc{0}{-q} + \ba{0}{q} \right), \nonumber 
\end{eqnarray}
where $H_{\text{C}}$ is defined in \Eq{eq:defH} and the last line comes from the intersubband part of the Coulomb interaction only (we have omitted the constant ground state energy).
If the $\ba{0}{q}$ operators are supposed to obey bosonic commutation rules
\begin{eqnarray}
\label{Comm}
\lbrack \ba{0}{q},\bc{0}{q'} \rbrack \simeq \delta_{\mathbf{q,q'}}\delta_{i,j},
\end{eqnarray}
then the Hamiltonian in \Eq{HIB} is quadratic and can thus be solved by an Hopfield-Bogoliubov transformation \cite{Hopfield58}.

The different approximations that led to \Eq{HIB} (parabolicity, small exchanged momentum, perfect bosonicity, absence of intrasubband scattering)
have been thorougly tested in multiple experiments \cite{Anappara07,Anappara09,Sapienza07,Sapienza08,Gunter09,Delteil11} in the linear regime. 

Still, these simplifying approximations oblige us to neglect a certain number of phenomena, like 
the interplay between nonparabolicity and Coulomb interaction \cite{Nikonov97}, or the scattering toward electron-hole pairs at large wavevectors, that is known to be an important factor in the thermalization and dynamics of exciton polaritons \cite{Porras02,Ciuti04}. 
Moreover neglecting intrasubband Coulomb scattering does not allow us to consider
polariton-polariton Coulomb interactions, that would appear as an higher order term in the Hamiltonian of \Eq{HIB}. 
Indeed, the bosonic approximation in \Eq{Comm} is satisfied only if its expectation value is taken between states corresponding to a weak excitation density. 
Otherwise, we will have a correction term in \Eq{Comm}, roughly proportional
to the ratio between the number of excitations, $\Nx$, and of electrons, $\Nel n_{\text{QW}}$, in such a way that this deviation can become important at higher excitation densities \cite{DeLiberato09}.

To go beyond these limitations, we will derive an effective quartic bosonic Hamiltonian, taking care of both nonbosonicity and Coulomb scattering. In order to do so, we will have to calculate the matrix elements of the full Hamiltonian in \Eq{HF} between states with multiple intersubband excitations.
The algebra necessary to calculate the coefficients of such an effective Hamiltonian is rather heavy, and we will use  the coboson formalism, originally developed for exciton polaritons \cite{Combescot07, Combescot08}.
The reader can refer to  the next section for a brief summary of such a formalism, highlighting the aspects relevant for the algebra of the present paper.

%%%%%%%%%%%%%%%%%%%%%%%%%%
\section{Composite boson commutator formalism}
\label{coboson}
Here we generalize the composite boson (coboson) approach, which was originally developed in Refs. [\onlinecite{Combescot07}] and [\onlinecite{Combescot08}] for the case of excitons. Such
a formalism applied to intersubband excitations provides a systematic way to calculate many-body matrix elements.
For simplicity, we consider the case where $\nqw=1$. The case with many quantum wells is more cumbersome, but relatively straightforward. Final results are obtained by replacing $1/\Nel$ by $1/(\nqw\Nel)$ in \Eq{eq:pauliDef}, (\ref{eq:twoISB}) and (\ref{eq:direct}).

We start by writing Hamiltonian $H_F$ in \Eq{HF} in the appropriate form to use the composite boson commutator formalism. In particular it must annihilate the ground state. Our main interest will be to calculate matrix elements describing scattering events between intersubband excitations. We will thus limit ourselves to consider matrix elements of the form $\gsbra \dots \ba{i}{q} H \bc{i'}{q'} \dots \gsket$, where all intersubband excitation creation operators are on the right of the Hamiltonian and annihilation operators on the left.
It is thus immediate to verify that terms in the Hamiltonian containing operators $\cc{1}{k}$ or $\ca{1}{k}$ in normal order, with $\vect{k}$ outside the Fermi sea, do not contribute to such matrix elements, as they annihilate the state.
We can thus remove them by truncating sums over $\vect{k}$, relative to the first subband, to the Fermi sea ($k\leq k_F$).
We then rewrite $H_{F}$ in  \Eq{HF} in the following form (omitting the ground state energy and non-contributing terms)
\begin{eqnarray}
\label{HFpre}
H_{F}&=&\tilde{H}_{\text{Kin}} +H_{\text{Rabi}} +H_{\text{Intra}}+H_{\text{Depol}} ,
\end{eqnarray}
where
\begin{eqnarray}
\tilde{H}_{\text{Kin}} & = & \sumk{k} \hbar\tilde{\omega}_{2,k} \cc{2}{k}\ca{2}{k} - \sumk{k} \hbar\tilde{\omega}_{1,k} \ca{1}{k}\cc{1}{k}  ,
\end{eqnarray}
\begin{eqnarray}
         H_{\text{Intra}} & = & \frac{1}{2} \sumk{k,k',q} \tilde{V}_{q}^{2222} \cc{2}{k+q}\cc{2}{k'-q}\ca{2}{k'}\ca{2}{k}  \nonumber  \\
         & & + \frac{1}{2} \sumk{k,k',q} \tilde{V}_{q}^{1111} \ca{1}{k+q}\ca{1}{k'-q}\cc{1}{k'}\cc{1}{k}  \nonumber  \\
         & & - \sumk{k,k',q} \tilde{V}_{q}^{1221} \ca{1}{k+q}\cc{2}{k'-q}\ca{2}{k'}\cc{1}{k}  ,
\end{eqnarray}
\begin{multline}
\label{eq:HDEP}
         H_{\text{Depol}}  =  \frac{\Nel}{2} \sumk{q} \tilde{V}_{q}^{1212} \left( 2\bc{0}{q}\ba{0}{q} \right. \\
         \left. + \bc{0}{q}\bc{0}{-q} + \ba{0}{q}\ba{0}{-q} \right),
\end{multline}	
all sums over the first subband are again truncated to the Fermi wavevector and	
\begin{eqnarray}
\hbar\tilde{\omega}_{1,k} & = & \hbar\omega_{1,k} - \sum_{|\vect{k}'|<k_{F}} \tilde{V}_{|\vect{k-k'}|}^{1111}\\
\hbar\tilde{\omega}_{2,k} & = &\hbar\omega_{2,k} - \sum_{|\vect{k}'|<k_{F}} \tilde{V}_{|\vect{k-k'}|}^{1212}\nonumber.
\end{eqnarray}	
In the previous equation we have thus explicitly separated the intersubband and intrasubband terms of the Coulomb interaction, putting them respectively into 
$H_{\text{Depol}}$ and $H_{\text{Intra}}$. The first term will give the depolarization shift \cite{DeLiberato12}, while the second
one contains the electron-hole interaction which lowers the energy of bare intersubband excitations \cite{Shelykh12} and is also responsible for the scattering between intersubband excitations.
Moreover, we commuted the electronic operators in such a way that now $\tilde{H}_{\text{Kin}}$ and $H_{\text{Intra}}$ annihilate the ground state $\ket{F}$. 
Writing the Hamiltonian in such a form makes it also apparent the renormalization effect of the Coulomb interaction over the electronic dispersions,
that is the same as the one that would be given by an Hartree-Fock approach\cite{Giuliani05, Shelykh12}.
 
The starting point of the composite boson formalism is the exact commutator between intersubband transitions defined in \Eq{isbt}
\begin{eqnarray}
	\commute{ \ba{m}{q}}{ \bc{i}{q'}}	&=&
	\delta_{m,i} \delta_{\mathbf{q} , \mathbf{q'}}   - D_{m\mathbf{q} , i\mathbf{q}'} ,
\end{eqnarray}
where $D_{m\mathbf{q} , i\mathbf{q}'} = D^{(1)}+D^{(2)}$ is the deviation from bosonicity operator defined by
\begin{eqnarray}
	D^{(1)}&=&
	\frac{1}{\Nel} \sumk{k} \nuc{m}{k} \nuc{i}{k+q-q'}^{\ast} \ca{1}{k+q-q'} \cc{1}{k} \nonumber  \\
	\label{eq:deviationHole}
	D^{(2)}&=&
	\frac{1}{\Nel} \sumk{k} \nuc{m}{k} \nuc{i}{k}^{\ast} \cc{2}{k+q'} \ca{2}{k+q} .
	\label{eq:deviationElec}
\end{eqnarray}
To calculate many-body matrix elements involving intersubband excitations, we need to evaluate the following commutator
\begin{eqnarray}
\label{eq:pauli}
	\commute{D_{m\mathbf{q} , i\mathbf{q}'}}{\bc{j}{q''}} &=&
	\frac{1}{\Nel} \sum_{n}
	\left\{  \pauli{n}{j}{m}{i}{q-q'} \right.   \\
	&& + \left. \pauli{n}{i}{m}{j}{q-q''}  \right\}
	\bc{n}{q'+q''-q},  \nonumber
\end{eqnarray}
where
\begin{eqnarray}
\label{eq:pauliDef}
	\pauli{n}{j}{m}{i}{q} &=&
	 \frac{1}{\Nel}  \sumk{k}  \nuc{m}{k} \nuc{n}{k+q} \nuc{i}{k+q}^{\ast} \nuc{j}{k}^{\ast}. 
\end{eqnarray}
The term appearing in \Eq{eq:pauli} is analogous to an effective scattering due to the nonbosonicity 
of the intersubband excitations (Pauli scattering): two intersubband excitations can exchange their electrons to create 
two excitations with different wavevectors, the total wavevector being conserved.

With the identities above mentioned, it is possible to compute the scalar products of many-excitation states as
\begin{eqnarray}
\label{eq:twoISB}
	\bra{F}  \ba{n}{q'''}\ba{m}{q}\bc{i}{q'}\bc{j}{q''}  \ket{F} = \delta_{m\vect{q},i\vect{q}'} \delta_{n\vect{q}''',j\vect{q}''} \\
	- \frac{1}{\Nel} \pauli{n}{j}{m}{i}{q-q'} + (m,\vect{q} \leftrightarrow n,\vect{q}''') \nonumber .
\end{eqnarray}
From Eqs. (\ref{eq:Hrabi}) and (\ref{eq:HDEP}) we see that matrix elements of $H_{\text{Rabi}}$ and $H_{\text{Depol}}$ can be written as scalar products of many-excitation states and can thus be calculated with the previous formula.

In order to evaluate the many-body matrix elements of the kinetic part of the Hamiltonian $\tilde{H}_{\text{Kin}}$, 
we need the commutator 
\begin{equation}
\label{eq:kinetic}
	\commute{\tilde{H}_{\text{Kin}}}{\bc{i}{q}} =
	\hbar \tilde{\omega}_{12} \bc{i}{q},
\end{equation}
where we have approximated, for simplicity, $\tilde{\omega}_{2,\lvert \mathbf{k+q}\lvert}$ with  $\tilde{\omega}_{2,k}$.
Many body matrix elements of $\tilde{H}_{\text{Kin}}$ can thus be calculated by commuting it all the way to the right, leaving behind only terms in the form of scalar products between states with many intersubband excitations.

To complete the framework, we need to consider the matrix elements of $H_{\text{Intra}}$. We start again by calculating its commutator with an intersubband excitation creation operator
\begin{eqnarray}
	\commute{H_{\text{Intra}} }{\bc{i}{q}} & = &
	- \sum_{j} \gamma_{i,j} \bc{j}{q} + V_{i, \mathbf{q}} ,
	\label{eq:defCreation}
\end{eqnarray} 
where $\gamma_{i,j}$ is a coupling between bright and dark excitations
\begin{eqnarray}
	\gamma_{i,j} & = &
	\sumk{q} \tilde{V}_{q}^{1221} \pauli{0}{0}{j}{i}{q} ,
	\label{eq:gamma}
\end{eqnarray}
and $V_{i,\vect{q}}$ is called the creation potential operator,
\begin{eqnarray}
	\label{eq:creationPotential}
	V_{i, \vect{q}} &=& \sum_{\vect{Q},m}   \Big(  \tilde{V}_{Q}^{2222} \delta_{m,i} - \tilde{V}_{Q}^{1221} \pauli{0}{0}{m}{i}{Q} \Big)  \\
	&&   \quad  \times  \bc{m}{q+Q}  \sum_{\vect{k}} \cc{2}{k-Q}\ca{2}{k}  \notag  \\
	& + & \sum_{\vect{Q},m}  \Big(  \tilde{V}_{Q}^{1111} \pauli{0}{0}{m}{i}{Q} - \tilde{V}_{Q}^{1221} \delta_{m,i} \Big)  \notag \\
	&&   \quad  \times \bc{m}{q+Q}  \sum_{\vect{k}} \ca{1}{k+Q}\cc{1}{k} .  \notag
\end{eqnarray}
The coefficient $\gamma_{i,i}$ in \Eq{eq:defCreation} is a renormalization of the intersubband transition energy due to the electron-hole interaction. Notice that the creation potential operator, $V_{i,\vect{q}}$, as well as $\tilde{H}_{\text{Kin}}$ and $H_{\text{Intra}}$, annihilate the ground state $\ket{F}$, a property which will be important when calculating matrix elements.
The last commutator we need is the one between a creation potential operator and an intersubband excitation creation operator, that is 
\begin{eqnarray}
\label{eq:direct}
	\commute{ V_{i, \vect{q}} }{ \bc{j}{q'} } & = &
	\frac{1}{\Nel} \frac{e^{2} \nel}{2\epsilon_{0}\epsilon_{r}} \\
	&& \sum_{m,n,\mathbf{Q}}
	\frac{ \direct{n}{j}{m}{i}(\mathbf{Q}) }{Q \epsilon(\vect{Q})} \,
	\bc{m}{q+Q} \bc{n}{q'-Q} , \nonumber \qquad
\end{eqnarray}
where
\begin{eqnarray}
	\direct{n}{j}{m}{i}(\vect{Q}) & = &
	\delta_{m,i} \delta_{n,j} I_{Q}^{2222} + \pauli{0}{0}{m}{i}{Q} \pauli{n}{j}{0}{0}{Q}  I_{Q}^{1111}  \nonumber \\
	&-& \delta_{m,i} \pauli{n}{j}{0}{0}{Q} I_{Q}^{1221} - \delta_{n,j} \pauli{0}{0}{m}{i}{Q} I_{Q}^{1221} ,  \nonumber
\end{eqnarray}
is the so-called direct scattering.

In order to calculate an arbitrary matrix element concerning $H_{\text{Intra}}$ between many-excitations states, we can commute it all the way to the right, leaving behind terms that will have the form of a creation potential operator, $V_{i, \vect{q}}$, sandwiched between multiple intersubband excitation operators. We can then commute creation potential operators all the way to the right, leaving at the end only many-excitations states scalar products, that we can calculate using \Eq{eq:twoISB}.

%%%%%%%%%%%%%%%%   
\section{Effective Bosonic Hamiltonian}
\label{effective}

Our aim is to derive a bosonic Hamiltonian capable of describing intersubband polariton-polariton interactions.
We will thus introduce the bosonic ground state $\ket{G}$ and $\Ba{q}$, the annihilation operator for a bright bosonic intersubband excitation, with in-plane wavevector $\mathbf{q}$ satisfying the usual bosonic commutation rules
\begin{eqnarray}
	\commute{ \Ba{q} }{ \Bc{q'} } & = &
	\delta_{\mathbf{q},\mathbf{q'}}  .
\end{eqnarray}
The index has been omitted as we are only interested in the dynamics of bright excitations.
We will then calculate the coefficients of the bosonic Hamiltonian $H_B$, imposing that it has the same expectation values that $H_F$, for states with up to two excitations.
While the most general Hamiltonian, up to $n$-body interactions, can be calculated with the same procedure, here we will limit ourselves to consider only resonant two-body interactions involving bright intersubband excitations.

The actual calculation of the coefficients requires some rather complex operatorial algebra. 
For sake of simplicity, we have thus moved the bulk of the calculations in the Appendices, presenting here only the final form of the Hamiltonian and the parameter dependency of its coefficients. 
In Sec. \ref{coboson}, we have provided the essential ingredients of the composite boson commutator approach \cite{Combescot07,Combescot08,DeLiberato09},
that we then use extensively in Appendix \ref{sec:coeff} to calculate the actual coefficients of the bosonic Hamiltonian. 

The bosonic Hamiltonian, describing the bright excitations, and using the quantities defined in Sec.~\ref{coboson} and in Appendix~\ref{sec:coeff}, has thus the  form
\begin{widetext}
\begin{eqnarray}
	H_{B} & = &
	\label{HB}
	\sumk{q}  \left( \hbar\tilde{\omega}_{12} - \gamma_{0,0} \right)   \, \Bc{q}\Ba{q}
	\, + \sumk{q}  \frac{e^{2} \nel}{4\epsilon_{0}\epsilon_{r}\epsilon_{12}(0)}  \left. \frac{I_{q}^{1212}}{q} \right|_{q=0}   \left[ 2\Bc{q}\Ba{q} + (1-\zeta) \left( \Ba{q}\Ba{-q}  +  \Bc{q}\Bc{-q} \right) \right]  \nonumber  \\
	&& + \sumk{q} \hbar\Omega_{q} \, (\Ba{q}+\Bc{-q})(\pa{-q}+\pc{q})  \nonumber \\
	&& - \frac{1}{\nqw \Nel}  \sumk{q,q',p}   \frac{\hbar\Omega_{q}}{2}  \left[  1 + \left( 1-\delta_{\mathbf{p,0}} \right) \left( 1-\delta_{\mathbf{p,q'-q}} \right)  \right]     \, \Bc{q+p}\Bc{q'-p}\Ba{q'}\pa{q}+ \text{h.c.}   \nonumber  \\
	&& - \frac{1}{\nqw \Nel}  \frac{1}{2}   \sumk{q,q',p}   U   \, \Bc{q+p}\Bc{q'-p}\Ba{q'}\Ba{q}  ,
\end{eqnarray}
\end{widetext}
where the last nonlinear coefficient, describing the scattering of pairs of intersubband excitations, is
\begin{eqnarray}
	U & = &
	\frac{e^{2} \nel}{\epsilon_{0}\epsilon_{r}\epsilon_{12}(0)}  \left. \frac{I_{q}^{1212}}{q} \right|_{q=0}   (1 - \zeta)  \,\,
	- \,\, \frac{e^{2} \nel}{2\epsilon_{0}\epsilon_{r}\kappa}  \direct{0}{0}{0}{0}(0) \nonumber \\&&
	+ \,\, \frac{e^{2} }{2\epsilon_{0}\epsilon_{r}}  \sqrt{\frac{\nel}{2\pi}} x(0) .
	 \label{eq:U}
\end{eqnarray}

In the previous expressions $\nel$ denotes the electron density in each quantum well.
A comparison with \Eq{HIB} shows that taking into account the nonbosonicity of intersubband excitations and the intrasubband terms of the Coulomb interaction resulted in the appearance of quartic terms in the Hamiltonian, that makes it possible to observe nonlinear polariton-polariton processes.
In the first and fourth line in \Eq{HB} and in \Eq{eq:U} coefficients have no dependence over the wavevectors of the intersubband excitations. This is justified by the fact that these coefficients evolve significantly on a scale of the order of the Fermi wavevector $k_{F}$, which is much larger than the photonic wavevectors of interest. There is also no spin dependence as one should expect from a naive comparison with excitons \cite{Ciuti98,Vladimirova10}. As explained in Sec.~\ref{sec:ISB} and Appendix~\ref{sec:notation} this is due to the fact that all intersubband of interest are made of two fermions with the same spin.

The first quartic term in \Eq{HB} describes the saturation of the light-matter interaction. Its physical origin lies in the non-perfect bosonicity of the intersubband excitations and accounts for a renormalization of the Rabi frequency at high excitation densities. Its presence is easy to understand if one remembers \cite{DeLiberato08} that the light-matter interaction is proportional to  the square root of the total electronic population in the first subband minus the one in the second subband. As the creation of one intersubband excitation corresponds, in average, to the promotion of one electron from the first to the second subband, we do expect that an increase in the intersubband excitations population will reduce the light-matter coupling, consistently with the minus sign in front of such a coefficient.

The second quartic term describes instead the scattering between pairs of intersubband excitations. It has three contributions as seen in \Eq{eq:U}. 
The first one comes from the intersubband Coulomb interaction.
The second and third contributions instead are due to the intrasubband Coulomb interaction.
It is important to notice that the two nonlinear coefficients behave differently with respect to the number of quantum wells in the cavity. While $\hbar\Omega_q$ is proportional to the square root of the total number of electrons in the cavity, $U$ depends only on the number of electrons in each well. This is due the fact that the photonic mode couples to all electrons in the cavity while the Coulomb interaction does not couple one well with another.

As we can see, Coulomb interaction and the non-perfect bosonicity also give a renormalization effect visible in the quadratic part of the Hamiltonian. 
The intersubband energy is renormalized both by Hartree-Fock terms and by the Coulomb electron-hole interaction. Moreover, the non-resonant terms of the depolarization shift are renormalized by a  $\zeta < 1$ coefficient as a consequence of the nonbosonicity.
Again, the expressions of the coefficients can be found in Appendix~\ref{sec:coeff}. 

The previous Hamiltonian is adapted to describe optically pumped experiments, in which the populations of the dark states remain negligible. While we will not treat this case in the present work, we notice that, in the case of electrical pumping, the dark states population can become significant. Interaction between bright and dark modes is described by the Hamiltonian
\begin{eqnarray}
\label{HD}
H_D&=&-2\sum_{i\neq 0,\mathbf{q,q',p}} \frac{G_{\mathbf{q-q',p}}^{i}}{\nqw \Nel} \, \Bc{q-p} B_{i,\mathbf{q'-p}}^{\dagger} B_{i,\mathbf{q'}} \pa{q} + \text{h.c.} \nonumber \\
&&-\sum_{i\neq 0,\mathbf{q,q',p}} \frac{U^{i}}{\nqw \Nel} \Bc{q+p} B_{i,\mathbf{q'-p}}^{\dagger} B_{i,\mathbf{q'}} \Ba{q},
\end{eqnarray}
where the index $i$ means that the mode is dark. In a mean field approach this Hamiltonian can be seen to describe energy shifts and saturation effects on the bright modes as a function of the dark state population.
Coefficients in \Eq{HD} can be calculated from a generalization of the formula given in Sec.~\ref{coboson} and in Appendix~\ref{sec:coeff}.

Passing from the fermionic description $H_{F}$ to the bosonic one $H_{B}$,
we have kept only quartic terms in $H_{B}$, discarding all higher order terms.
This is effectively equivalent to limit ourselves to the lowest order in the excitation fraction $\Nx / (\nqw \Nel)$  \cite{Combescot08}.
We thus expect that our theory will correctly describe the polariton-polariton interactions in the regime of moderate excitation densities\cite{Glazov09}.
In order to clarify this point we give here an explicit example not involving Coulomb interaction. The reader can refer to Appendix~\ref{sec:fermi} for more complex and detailed examples.

We consider a process in which a state with one photon and  $\Nx$ intersubband exitations is converted into a state with no photon and $\Nx+1$ intersubband excitations.
The matrix element of the Coulomb part of the Hamiltonian vanishes in this case, as it conserves the number of photons, and we can thus perform the calculation exactly, both in the fermionic and in the bosonic formalism.
We obtain respectively
\begin{widetext}
\begin{eqnarray}
	\frac{\bra{F} \pa{q} \ba{0}{q}^{\Nx}H_{F} b^{\dagger \, \Nx+1}_{0,\mathbf{q}}\ket{F}}{\sqrt{ \bra{F} \ba{0}{q}^{\Nx} b^{\dagger \, \Nx}_{0,\mathbf{q}}\ket{F} \bra{F} \ba{0}{q}^{\Nx+1} b^{\dagger \, \Nx+1}_{0,\mathbf{q}}\ket{F} }}   &=&   \hbar\Omega_{q} \, \sqrt{\Nx+1} \, \sqrt{1-\frac{\Nx}{\nqw \Nel}} \nonumber \\
	  &=&  \hbar\Omega_{q} \, \sqrt{\Nx+1} \left( 1-\frac{\Nx}{2\nqw\Nel} \right) + O\left( \frac{\Nx^2}{\nqw^2 \Nel^2} \right)   \\
	\frac{\bra{G} \pa{q} \Ba{q}^{\Nx}H_{B} B^{\dagger \, \Nx+1}_{\mathbf{q}}\ket{G}}{\sqrt{(\Nx+1)!\Nx!}}  &=& 
	\hbar\Omega_{q} \, \sqrt{\Nx+1} \left( 1-\frac{\Nx}{2\nqw\Nel} \right), \nonumber 
\end{eqnarray}
\end{widetext}
which are indeed equal to first order in $\Nx / (\nqw \Nel)$.

%%%%%%%%%%%%%%
\section{Numerical results}
\label{numerical}

In this Section we provide the numerical values of the relevant coefficients that appear in the bosonic Hamiltonian in \Eq{HB}, highlighting the dependence of the main parameters.

%%%%%%
\subsection{Renormalization of the intersubband transition}
\begin{figure}[t!]
\begin{center}
\includegraphics[width=9cm]{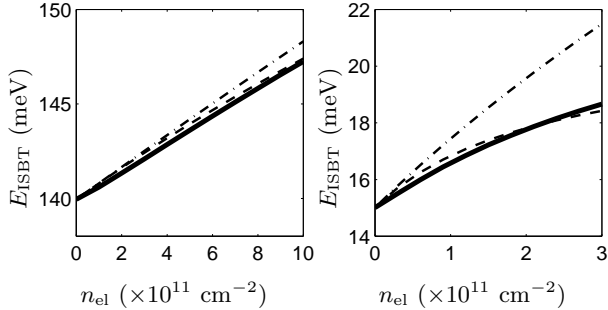}
\caption{Renormalized energy $E_{\text{ISBT}}$ of the intersubband  transition for a bare energy $\hbar\omega_{12}=140$meV (left panel) and $15$meV  (right panel), as a function of the electron density. 
Solid line: result including all the contributions  in \Eq{eq:EISBT}, namely Coulomb interaction and nonbosonicity. Dashed line: result neglecting intrasubband Coulomb interactions and the non-bosonicity. Dash-dotted line: same as the dashed line, but with no screening of the intersubband Coulomb interaction.}
\label{fig:plasmon}
\end{center}
\end{figure}

We start by analyzing the renormalization effect due to the Coulomb interaction on the quadratic part of the effective Hamiltonian. Performing a Bogoliubov transformation on the first line of \Eq{HB} we obtain the renormalized energy of the intersubband excitation (that we should call, more properly, intersubband plasmon \cite{Todorov10,Todorov12,DeLiberato12,Shelykh12})
\begin{widetext}
\begin{eqnarray}
	E_{\text{ISBT}} & = &
	\sqrt{ \left( \hbar\tilde{\omega}_{12} - \gamma_{0,0} + \frac{e^{2}\nel}{2\epsilon_{0}\epsilon_{r}\epsilon_{12}(0)} \left. \frac{I_{q}^{1212}}{q} \right|_{q=0} \right)^{2}  -  \left( \frac{e^{2}\nel}{2\epsilon_{0}\epsilon_{r}\epsilon_{12}(0)} \left. \frac{I_{q}^{1212}}{q} \right|_{q=0} (1-\zeta) \right)^{2} } .
	\label{eq:EISBT}
\end{eqnarray}
\end{widetext}
In Fig.~(\ref{fig:plasmon}) we plot the  dispersion of the intersubband transition energy considering a GaAs quantum well of length $L_{\text{QW}} = 11 \text{ nm}$ (left panel) and $L_{\text{QW}} = 39 \text{ nm}$ (right panel), corresponding to bare transitions of $140 \text{ meV}$ \cite{Dini03,Gunter09} and $15 \text{ meV}$ \cite{Todorov10} respectively. 
The solid line depicts the result for the intersubband transition including all contributions in Eq. (\ref{HB}).
The dashed line represents the result obtained while neglecting the Coulomb intrasubband interactions and the nonbosonicity.  The dash-dotted line is the result obtained as the dashed line, but neglecting also the screening of the intersubband Coulomb interaction. Notice that the renormalized intersubband energy, $E_{\text{ISBT}}$, in \Eq{eq:EISBT}, converges to the bare transition energy $\hbar\omega_{12}$ for vanishing doping.
We see that the main effect is due to the screening of the intersubband Coulomb interaction. It appears clear that these renormalization effects are relevant in the THz regime (right panel), but rather moderate for mid-infrared transitions (left panel).

%%%%%
% figure
\psfrag{x31}[tc]{$\nel \ (\times10^{11}\text{ cm}^{-2})$}
\psfrag{y31}[Bc]{$E_{\text{ISBT}} \text{ (meV)}$}
\psfrag{x32}[tc]{$\nel \ (\times10^{11}\text{ cm}^{-2})$}
\psfrag{y32}[Bc]{$E_{\text{ISBT}} \text{ (meV)}$}
%%%%%%%%

%%%%%
% figure
\psfrag{x41}[tc]{$\nel \ (\times10^{11}\text{ cm}^{-2})$}
\psfrag{y41}[Bc]{$\quad\quad U \text{ (meV)}$}
\psfrag{x42}[tc]{$\nel \ (\times10^{11}\text{ cm}^{-2})$}
\psfrag{y42}[Bc]{$\quad\quad U \text{ (meV)}$}
\begin{figure}[t!]
\begin{center}
\includegraphics[width=9cm]{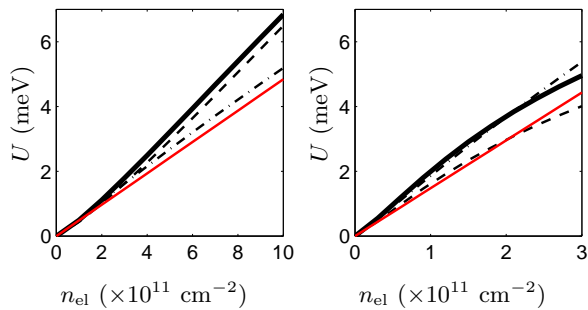}
\caption{Thick solid line: effective interaction energy $U$  between intersubband excitations including all the contributions  in \Eq{eq:U} for $\hbar\omega_{12}=140$meV (left panel) and $\hbar\omega_{12} = 15$meV (right panel) as a function of the electron density.
Dashed-line: the first term in  \Eq{eq:U} corresponding to the intersubband Coulomb interaction. Thin red line: the absolute value of the second term in  \Eq{eq:U}, namely the direct Coulomb interaction. 
Note that this term is negative, thus producing a red-shifted contribution. Dash-dotted line:  the third term in \Eq{eq:U}, which is due to the exchange Coulomb interaction.}
\label{fig:fourcoef}
\end{center}
\end{figure}
%%%%%%%%

%%%%%%
\subsection{Interaction energy between intersubband excitations}
\label{interaction}
Now, we consider the quartic part of the Hamiltonian $H_{B}$, responsible for the Coulomb scattering of pairs of intersubband excitations. The interaction energy is given by $\Nx U / (\nqw\Nel)$.
In Fig. \ref{fig:fourcoef} we plot the energy $U$ (thick solid line) for a mid-infrared transition (left panel) and a THz transition (right panel). 
The other lines depict the individual contributions of the three terms in \Eq{eq:U} (see caption for details). For the considered realistic parameters, the interaction energy grows with increasing electron doping density almost linearly for the mid-infrared case, while it saturates at high densities for the case of a THz transitions. Notice that in the case of excitons\cite{Ciuti98,Combescot08}, the direct scattering vanishes in the limit of small wavevectors, whereas it is not the case for intersubband excitations when the screened Coulomb interactions are considered. This is due the fact that electron-electron and hole-hole interactions are screened differently by the Fermi sea\cite{Lee99}.
Note that for a doping density in the range of a few $10^{11} {\rm cm}^{-2}$, the energy $U$ is of the order of a few meV both for the cases of THz and mid-infrared transitions. The interaction energy can thus reach values close to the meV when the  density of excitations in the system becomes significant.
This is rather promising, since as shown in the case of exciton-polaritons\cite{Carusotto12}, very interesting nonlinear polariton physics occurs when the interaction energy becomes comparable to the linewidth of the polariton modes.
For THz polaritons, state-of-the-art samples\cite{Todorov10} exhibits polariton width in the meV range.

%%%%%%%%%%%%%%%
\section{Polariton Hamiltonian}
\label{scattering}

In this Section we consider the interactions in the polariton basis.
For simplicity we limit ourselves to the first mode of the cavity and we assume that the number of electrons is such that the system is in the strong coupling regime. The diamagnetic interaction in $H_{C}$ as well as all non-resonant terms in $H_{B}$ are thus neglected.  A Bogoliubov transformation of the quadratic parts of $H_{C}$ and $H_{B}$ gives the expression of the polaritonic operators
\begin{eqnarray}
	\left(
	\begin{matrix}
		\lpa{q} \\ \upa{q}
	\end{matrix}  \right)
	& = &
	\left(
	\begin{matrix}
		\betaHopf{q} & -\alphaHopf{q} \\ \alphaHopf{q} & \betaHopf{q}
	\end{matrix} \right)
	\left(
	\begin{matrix}
		\Ba{q} \\ \pa{q}
	\end{matrix} \right)
	\label{eq:polaritons}
\end{eqnarray}
where $\upa{q}$ and $\lpa{q}$ are polaritonic operators of the upper (UP) and lower branch (LP) respectively, while $\alphaHopf{q}$ and $\betaHopf{q}$ are the Hopfield coefficients. Note that
results can be generalized to the ultrastrong coupling regime (vacuum Rabi frequency comparable to intersubband transitions) by considering the Bogoliubov transformation involving the antiresonant light-matter interaction
terms \cite{Ciuti05}. Here, for simplicity, we consider the case of moderate values of the vacuum Rabi frequency.
Because of the quartic terms in Hamiltonian $H_{B}$ polaritons interact with each other through their matter part and can scatter. In the following, we will focus on the lower branch. Replacing operators $\pa{q}$ and $\Ba{q}$ by their expression, the two-body interaction between LP polaritons is found to be
\begin{eqnarray}
	H_{\text{LP-LP}} & = &
	\frac{1}{2}  \sumk{q,q',p} \frac{V_{\mathbf{q,q',p}}}{\nqw \Nel} \,
	\lpc{q+p} \lpc{q'-p} \lpa{q'} \lpa{q} \,, \quad
	\label{eq:quarticPolariton}
\end{eqnarray}
where
\begin{eqnarray}
	V_{\mathbf{q,q',p}} & = &
	\betaHopf{q+p} \betaHopf{q'-p} \betaHopf{q'}
	\left(  \alphaHopf{q} \hbar\Omega_{q} - \betaHopf{q} U  \right) .
\end{eqnarray}

As an example, we now consider the case where the lower branch is pumped at a wavevector $\mathbf{q}_{\text{p}}$ so that the system is in the state $p_{\text{L}\,\mathbf{q}_{\text{p}}}^{\dagger \, \Nx} \ket{G} / \sqrt{\Nx !}$.
This Hamiltonian allows us to describe single-mode (Kerr) and multimode (parametric) coherent non-linearities\cite{Carusotto12}.
For the parametric case, one has to consider interaction channel
\begin{eqnarray}
	p_{\text{L}\,\mathbf{q}_{\text{p}}}^{\dagger \, \Nx} \ket{G} & \rightarrow &
	p_{\text{L}\,\mathbf{q}_{\text{p}}+\mathbf{p}}^{\dagger} p_{\text{L}\,\mathbf{q}_{\text{p}}-\mathbf{p}}^{\dagger} p_{\text{L}\,\mathbf{q}_{\text{p}}}^{\dagger \, \Nx-2} \ket{G} ,
	\label{eq:scattering}
\end{eqnarray}
where pairs of polaritons scatter from the pumped mode into signal-idler pairs. As for the case of exciton-polaritons we expect that the maximum efficiency of this parametric processes is achieved when the energy conservation condition is fulfilled\cite{Savvidis00}.
A mean-field approach of the problem\cite{Ciuti03,Carusotto12} shows that the matrix element between the initial and the final states is the relevant quantity to consider and has to be compared with the lifetime of the excitations.
For high pump intensity {i.e.} $\Nx \gg 1$ this matrix element is
\begin{eqnarray}
	M_{\mathbf{q}_{\text{p}},\mathbf{p}} & = &
	\frac{\Nx}{\nqw \Nel}  V_{\mathbf{q}_{\text{p}},\mathbf{q}_{\text{p}},\mathbf{p}} .
\end{eqnarray}
As discussed in Sec. \ref{interaction} and  shown in Fig. \ref{fig:fourcoef}, for THz polaritons nonlinear interaction energies of the order of a few meV (thus comparable to THz polariton linewidths) can be achieved, thus paving the way
to a very interesting coherent nonlinear physics for this kind of composite excitations.

%%%%%%%%%%%
\section{Conclusions}
\label{conclusions}

In this paper, we have presented comprehensively a microscopic theory for the manybody physics of intersubband cavity polaritons.
Using a composite boson commutator approach, we have derived the manybody interaction matrix elements involving (screened) Coulomb interaction processes
and Pauli saturation effects. We have also derived an effective bosonic Hamiltonian reproducing the matrix elements in the two-excitation manyfold calculated by the composite boson approach.
As a first application, we have calculated the renormalization of the intersubband transition frequency at the mean-field level and determined the strength of the interactions between intersubband excitations.
We have also derived the renormalization of the light-matter coupling due to the Pauli saturation effects and considered the interaction physics in the polariton basis.
Our results show that significant intersubband polariton-polariton interactions occur, especially for transitions in the THz regime.
Our work paves the way to promising future studies of nonlinear quantum optics with intersubband cavity polaritons.

\acknowledgements We acknowledge partial support from the french ANR project THINQE-PINQE. MB is also supported by KAKENHI 24-632. We wish to thank C. Sirtori, Y. Todorov, A. Vasanelli for discussions.

%%%%%%%%%%%%%%%%%%%%%%%%%%%%%%%%%%%%%%
%%%%%%%   Appendix  %%%%%%%%%%%%%%%%%%%%%%%%
%%%%%%%%%%%%%%%%%%%%%%%%%%%%%%%%%%%%%%
\appendix

%%%%%%%%%%%%%%%%%%%%%%%%%%%%%%%%%%%%%%
\section{Spin and quantum well number}
\label{sec:notation}

In this appendix we give the explicit notations for \Eq{eq:HFdetail} and \Eq{eq:isb0} with quantum well and spin indexes.
Electronic wavefunctions are localized in quantum wells and we neglect electronic tunneling from one well to another. We thus neglect Coulomb interaction between electrons in different wells, which are sufficiently apart. \Eq{eq:HFdetail} and~(\ref{eq:isb0}) can then be written with all indexes
\begin{widetext}
\begin{eqnarray}
\label{eq:HFdetailIndex}
	H_{\text{Kin}} & = & \sum_{j=1}^{\nqw} \sum_{\mathbf{k},\sigma, \mu }
	\hbar\omega_{\mu,k} \,
	c_{\mu,\mathbf{k},\sigma}^{(j)\, \dagger} \, c_{\mu,\mathbf{k},\sigma}^{(j)}  \\
	H_{\text{Rabi}} & = & \sum_{j=1}^{\nqw} \sum_{\mathbf{k},\mathbf{q},\sigma} \hbar\chi_q \,
	(c_{2,\mathbf{k+q},\sigma}^{(j) \, \dagger} \, c_{1,\mathbf{k},\sigma}^{(j)} +
	c_{1,\mathbf{k+q},\sigma}^{(j) \, \dagger} \, c_{2,\mathbf{k},\sigma}^{(j)}) \, (\pa{q}+\pc{-q}) \nonumber \\
	H_{\text{Coul}} & = & \frac{1}{2} \sum_{j=1}^{\nqw} \sum_{ \substack{\mathbf{k, k' ,q},\sigma,\sigma' \\ \mu, \mu' , \nu, \nu'} }
	V_{ q }^{ \mu \nu \nu' \mu' } \,
	c_{\mu,\mathbf{k+q},\sigma}^{(j) \, \dagger} \, c_{\nu,\mathbf{k'-q},\sigma'}^{(j) \, \dagger} \, c_{\nu',\mathbf{k'},\sigma'}^{(j)} \, c_{\mu',\mathbf{k},\sigma}^{(j)} , \nonumber
\end{eqnarray}
\end{widetext}
and
\begin{eqnarray}
\label{eq:isb0Index}
	\bc{0}{q} & = &
	\frac{1}{ \sqrt{\nqw \Nel} }
	\sum_{j=1}^{\nqw} \sum_{\mathbf{k},\sigma} \nu_{0,j,\mathbf{k}}^{\ast} \,
	c_{2,\mathbf{k+q},\sigma}^{(j) \, \dagger} \,  c_{1,\mathbf{k},\sigma}^{(j)}  ,
\end{eqnarray}
where $\nu_{0,j,\mathbf{k}} = \Theta(k_F-k)$ for all $j$, $\mathbf{k}$ is a wavevector such that $k<k_F$, $\sigma,\sigma' \in \{\downarrow,\uparrow\}$ and $\mu, \mu', \nu, \nu' \in \{1,2\}$.
Bright intersubband excitations are thus linear superposition of pairs of fermions with the same spin. We could generalize \Eq{eq:isb0Index} by allowing the two fermions to have different spins but the resulting collective excitation would be dark and thus not relevant if we consider only polariton.

%%%%%%%%%%%%%%%%%%%%%%%%%%%%%%%%%%%%%%
\section{Coefficient calculation}
\label{sec:coeff}

In order to determine the coefficients of the effective bosonic Hamiltonian in \Eq{HB}, we impose that it has the same
matrix elements as the fermionic one in \Eq{HF} over the Hilbert space spanned by one and two excitation states.
That is we impose that
\begin{eqnarray}
\label{constr}
\frac{ \bra{G}\Ta{q}\Ta{q'} H_B \Tc{q''}\Tc{q'''}\ket{G} }{ \sqrt{ N_{\mathbf{q,q'}} N_{\mathbf{q'',q'''}} } } & = &
\frac{ \bra{F}\ta{q}\ta{q'} P_{\bot} H_F \tc{q''}\tc{q'''}\ket{F} }{ \sqrt{ A_{\mathbf{q,q'}} A_{\mathbf{q'',q'''}} } } \nonumber ,\\
\end{eqnarray}
where $\Ta{q}\in \lbrack \mathbb{I}, \pa{q},\Ba{q}\rbrack$, $\ta{q}\in \lbrack  \mathbb{I},\pa{q},\ba{0}{q}\rbrack$, $P_{\bot}$ is the projector over the subspace orthogonal to the ket\cite{Glazov09} and $N_{\mathbf{q,q'}}$ and $A_{\mathbf{q,q'}}$ are the normalizations of the orthogonalized states.
 
The constraints in \Eq{constr}, together with the fact that we decided to consider, beyond quadratic terms,
only the quartic ones, are enough to uniquely define $H_{B}$. The calculation of the coefficients of
$H_{B}$ thus reduces to the calculation of  matrix elements of the fermionic Hamiltonian $H_F$, that we can perform 
exploiting the commutator approach outlined in Sec. \ref{coboson}.

We thus start from a general bosonic Hamiltonian
\begin{eqnarray}
\label{HBtest}
H_B &=& \sumk{q} K \, \Bc{q}\Ba{0}{q} + \sumk{q}  Q \, \Ba{q}\Ba{-q}  + \text{h.c}  \nonumber \\
&+& \sumk{q} \hbar\tilde{\Omega}_{q} (\Ba{q}+\Bc{-q})(\pa{-q}+\pc{q})  \nonumber \\
&-&\sumk{q,q',p}  \frac{G_{\mathbf{q-q',p}}}{\nqw \Nel} \Bc{q+p}\Bc{q'-p}\Ba{q'}\pa{q}+ \text{h.c.}  \nonumber \\
&-&\frac{1}{2} \sumk{q,q',p}  \frac{U}{\nqw \Nel} \Bc{q+p}\Bc{q'-p}\Ba{q'}\Ba{q}.
\end{eqnarray}
We also introduce the following five matrix elements (each one implicitly dependent on all the relevant wavevectors)
\begin{eqnarray}
\label{dcoeff}
	d_{1} & = & \gsbra \ba{0}{q} H_F \bc{0}{q} \gsket   \nonumber    \\
	d_{2} & = & \gsbra \ba{0}{q} H_F \pc{q} \gsket      \nonumber  \\
	d_{3} & = & \gsbra H_F \bc{0}{q} \bc{0}{-q} \gsket    \nonumber \\
	d_{4} & = & \gsbra \ba{0}{q+p}\ba{0}{q'-p}H_F \bc{0}{q'}\pc{q} \gsket    \nonumber \\
	d_{5} & = & \gsbra \ba{0}{q+p}\ba{0}{q'-p} P_{\bot} H_F \bc{0}{q'}\bc{0}{q} \gsket  .
\end{eqnarray}
The relations between the coefficients of $H_B$ in \Eq{HBtest} and the matrix elements in \Eq{dcoeff} can be found with some algebra to be
 \begin{eqnarray}
	K & = & d_{1}\nonumber \\
	2 Q & = & \frac{d_{3}}{\sqrt{1-2/\nqw \Nel}} \nonumber \\
	\hbar\tilde{\Omega}_{q} & = & d_{2}  \nonumber
\end{eqnarray}
\begin{multline}
	\left( \delta_{\mathbf{p},0} + \delta_{\mathbf{q'-q,p}} \right) \hbar\tilde{\Omega}_{q} - \frac{G_{\mathbf{q-q',p}}}{\nqw \Nel} - \frac{G_{\mathbf{q-q',q'-q-p}}}{\nqw \Nel}      \\
	 = \sqrt{ \frac{ N_{\mathbf{q+p,q'-p}} }{ A_{\mathbf{q+p,q'-p}} } } \, d_{4}
\end{multline}
and
\begin{multline}
	2 \left( \delta_{\mathbf{p},0} + \delta_{\mathbf{q'-q,p}} \right) K - \frac{2 U}{\nqw \Nel}  \\
	 = \sqrt{ \frac{ N_{\mathbf{q,q'}} N_{\mathbf{q+p,q'-p}} }{ A_{\mathbf{q,q'}} A_{\mathbf{q+p,q'-p}} } } \, d_{5}  ,
\end{multline}
where, for consistency, terms under the square root have also to be developed to the first order in $1/(\nqw\Nel)$.

We have thus reduced the determination of the coefficients of $H_{B}$ to the calculation of the five matrix elements in \Eq{dcoeff}.
 Without loss of generality, in the following calculations,  we will assume that the excitations wavevectors are small compared to the Fermi wavevector,
which implies that the Pauli scattering involving two bright excitations, defined in \Eq{eq:pauliDef}, can be approximated by a Kroneker delta
\begin{eqnarray}
	\pauli{0}{0}{i}{j}{p} & \approx & \delta_{i,j}
	\label{eq:approximation}.
\end{eqnarray}

%%%%%%
\subsection{Coefficient $d_1$}

The only contributing terms are the kinetic and Coulomb part of the Hamiltonian $H_{F}$. The matrix element is
\begin{eqnarray}
	d_{1} &=& \hbar\tilde{\omega}_{12} - \gamma_{0,0}
	+ \frac{e^{2} \nel}{2\epsilon_{0}\epsilon_{r}\epsilon_{12}(0)}
	\left. \frac{I_{q}^{1212}}{q} \right|_{q=0} .
\end{eqnarray}
%which has two contributions. The first one si the free electron-hole energy lowered by the electron-hole Coulomb interaction while the second one is the blueshift due to the depolarization term. 

%%%%%%
\subsection{Coefficient $d_2$}

Only the light-matter part of $H_{F}$ contributes and we obtain
\begin{equation}
	d_{2} = \hbar \Omega_{q} .
\end{equation}

%%%%%%
\subsection{Coefficient $d_3$}

This term comes from the intersubband Coulomb interaction and is given by
\begin{eqnarray}
	d_{3} & = &
	\frac{e^{2} \nel}{2\epsilon_{0}\epsilon_{r}\epsilon_{12}(0)}
	\left. \frac{I_{q}^{1212}}{q} \right|_{q=0}
	\left( 1 - \zeta \right),
\end{eqnarray}
where
\begin{eqnarray}
	\left. \frac{I_{q}^{1212}}{q} \right|_{q=0} \times \zeta & = &
	\frac{1}{\Nel} \sumk{Q} \frac{I_{Q}^{1212}}{Q} \pauli{0}{0}{0}{0}{Q} .
\end{eqnarray}
The coefficient $Q$ is then $d_{3} (1 + 1/\nqw\Nel)$. As the number of electrons is large we can neglect the correction due to the normalization.

%%%%%%
\subsection{Coefficient $d_4$}

As for the coefficient $d_{2}$, only the light-matter part of $H_{F}$ contributes, giving
\begin{equation}
	d_{4} = \hbar \Omega_{q}
	\left(  \delta_{\mathbf{p},0} + \delta_{\mathbf{p,q'-q}} - \frac{2}{\nqw \Nel}  \right).
	\label{eq:d4}
\end{equation}
The last term, at the origin of the first quartic term in \Eq{HB}, is due to the nonbosonicity.

%%%%%%
\subsection{Coefficient $d_5$}

The expression of the coefficient is given by the following relations (to first order in $1/(\nqw\Nel$))
\begin{widetext}
\begin{eqnarray}
	d_{5} & = &
	\left(  \gsbra \ba{0}{q+p}\ba{0}{q'-p} - \gsbra \ba{0}{q+p}\ba{0}{q'-p} \bc{0}{q'}\bc{0}{q} \gsket \gsbra \ba{0}{q}\ba{0}{q'}  \right)
	H_{F} \bc{0}{q'}\bc{0}{q} \gsket ,
\end{eqnarray}
\begin{eqnarray}
	\gsbra \ba{0}{q+p}\ba{0}{q'-p} H_{F} \bc{0}{q'}\bc{0}{q} \gsket & = &
	2 \left( \hbar\tilde{\omega}_{12} - \gamma_{0,0} \right) \left( \delta_{\mathbf{p},0} + \delta_{\mathbf{p,q'-q}} - \frac{2}{\nqw\Nel} \right)  \nonumber \\
	&& + \frac{e^{2}\nel}{\epsilon_{0}\epsilon_{r}\epsilon_{12}}  \left. \frac{I_{q}^{1212}}{q} \right|_{q=0}   \left( \delta_{\mathbf{p},0} + \delta_{\mathbf{p,q'-q}} - \frac{4-2\zeta}{\nqw\Nel} \right)  \nonumber \\
	&& + \frac{2}{\nqw\Nel} \left( \frac{e^{2}\nel}{2\epsilon_{0}\epsilon_{r}\kappa} \direct{0}{0}{0}{0}(0) - \frac{e^{2}}{2\epsilon_{0}\epsilon_{r}} \sqrt{ \frac{\nel}{2\pi} }  x(0) \right) ,
	\label{eq:matrixGeneral}
\end{eqnarray}
\end{widetext}
where
\begin{eqnarray}
	x(\mathbf{p}) & = & \sum_{m,n,\mathbf{Q}}  \frac{k_{F}}{\Nel Q \epsilon(\vect{Q})}  \pauli{0}{n}{0}{m}{Q-p}
	\, \direct{n}{0}{m}{0}(-\vect{Q}) . \quad
	\label{eq:exchange}
\end{eqnarray}
In \Eq{eq:matrixGeneral} the long wavelength limit has been taken because the variation of the coefficients is significant only on a scale of the order of the Fermi wavevector $k_{F}$.

%\begin{widetext}
%$\clubsuit$
%To calculate $x_{\mathbf{p}}^{0,0}$ we use Motoaki's simplification
%\begin{eqnarray}
%	\sum_{n} \pauli{0}{n}{0}{0}{Q-p} \pauli{n}{0}{0}{0}{-Q} & = &
%	\frac{1}{\Nel} \sumk{k} \nuc{0}{k} \nuc{0}{k+Q-p} \nuc{0}{k+Q}^{\ast}   \nonumber  \\
%	\sum_{m,n} \pauli{0}{n}{0}{m}{Q-p} \pauli{0}{0}{m}{0}{-Q} \pauli{n}{0}{0}{0}{-Q} & = &
%	\frac{1}{\Nel} \sumk{k} \nuc{0}{k} \nuc{0}{k+Q-p} \nuc{0}{k-p}^{\ast} \nuc{0}{k+Q}^{\ast}
%\end{eqnarray}
%$\clubsuit$
%\end{widetext}

%%%%%%%%%%%%%%%%%%%%%%%%%%%%%%%%%%%%%%
\section{Transition rates and lifetimes}
\label{sec:fermi}

In this Appendix we calculate the transitions rates for scattering of pairs of intersubband excitations using both the effective bosonic Hamiltonian approach and the fermionic formalism developed by M. Combescot and coworkers \cite{Combescot07}, showing that the two approaches give exactly the same results

%%%%%%
\subsection{Fermionic case}

In this paragraph we use the composite boson commutator formalism to calculate the transition rate 
between an initial state $\ket{\psi_{i}}$ and a final state $\ket{\psi_{f}}$, as well as the 
lifetime of the former, using  a first-order, time-dependent perturbation theory. 
The calculation is equivalent to a Fermi golden rule as explained in Ref. [\onlinecite{Combescot07}].

The particular event we want to describe is the scattering of an initial pump beam of arbitrary intensity into a signal and an idler mode. We will thus consider initial and final states to be proportional respectively to the kets
\begin{eqnarray}
\ket{\psi_{i}}&\propto& b_{0,\mathbf{q}}^{\dagger \, \Nx} \gsket
\end{eqnarray}
and
\begin{eqnarray}
\ket{\psi_{f}}&\propto& \bc{0}{q+p} \bc{0}{q-p} b_{0,\mathbf{q}}^{\dagger \, \Nx-2} \gsket.\\\nonumber  
\end{eqnarray}

The transition rate can then be written as \cite{Combescot07}
\begin{eqnarray}
	t \Gamma_{\mathbf{p}} & = &  \left|  \langle \psi_{f} \big| \tilde{\psi}_{t} \rangle  \right|^{2} ,
\end{eqnarray}
where
\begin{eqnarray}
	\ket{\tilde{\psi}_{t}} & = & F_{t}(H_{F} - \bra{\psi_{i}}H_{F}\ket{\psi_{i}}) P_{\bot} H_{F} \ket{\psi_{i}} .
	\label{eq:psit}
\end{eqnarray}
As in Appendix~\ref{sec:coeff}, $P_{\bot}$ is the projector over the subspace orthogonal to $\ket{\psi_{i}}$ and $F_{t}$ verifies
\begin{eqnarray}
	|F_{t}(E)|^{2} & = &  \frac{2\pi t}{\hbar} \delta_{t}(E) ,
\end{eqnarray}
where $\delta_{t}$ converges to the Dirac delta function for long times.
The orthogonal projection in \Eq{eq:psit} can be calculated as
\begin{widetext}
\begin{eqnarray}
	P_{\bot} H b_{0,\mathbf{q}}^{\dagger \, \Nx} \gsket & = &
	- \Nx (\hbar\tilde{\omega}_{12}-\gamma_{0,0}) \sum_{i \neq 0} \gamma_{0,i} \bc{i}{q} b_{0,\mathbf{q}}^{\dagger \, \Nx-1} \gsket  +  \frac{\Nx (\Nx-1)}{2 \nqw\Nel} U b_{0,\mathbf{q}}^{\dagger \, \Nx} \gsket  \nonumber  \\
	&& + \frac{\Nx (\Nx-1)}{2 \nqw\Nel} \frac{e^{2}\nel}{2\epsilon_{0}\epsilon_{r}} \sum_{m,n,\mathbf{Q}} \frac{\direct{m}{0}{n}{0}(\mathbf{Q})}{Q \epsilon(\mathbf{Q})}  \bc{m}{q+Q} \bc{n}{q-Q} b_{0,\mathbf{q}}^{\dagger \, \Nx-2} \gsket  \nonumber  \\
	&& - \frac{\Nx (\Nx-1)}{2 \nqw\Nel}  \frac{e^{2}\nel}{\epsilon_{0}\epsilon_{r}\epsilon_{12}} \sum_{m,\mathbf{Q}} \pauli{m}{0}{0}{0}{Q-q} \bc{0}{Q} b_{m,2\mathbf{q-Q}}^{\dagger} b_{0,\mathbf{q}}^{\dagger \, \Nx-2} \gsket  .
\end{eqnarray}
\end{widetext}
The previous expression is a first order in the pertubations. We can thus consider only the zero-th order in $F_{t}$, {\itshape i.e} we do as if intersubband excitations were exact eigenstates of $H_{F}$, obtaining
\begin{widetext}
\begin{eqnarray}
	\gsbra \ba{0}{q}^{\Nx-2} \ba{0}{q-p} \ba{0}{q+p} F_{t}(H_{F} - \bra{\psi_{i}}H_{F}\ket{\psi_{i}}) & \approx & F_{t}(0) \gsbra \ba{0}{q}^{\Nx-2} \ba{0}{q-p} \ba{0}{q+p} .
\end{eqnarray}
\end{widetext}
Taking into account the normalization we thus obtain the transition rate
\begin{eqnarray}
	\Gamma_{\mathbf{p}} & = & \frac{2\pi}{\hbar} \frac{\Nx (\Nx-1)}{\nqw^{2}\Nel^{2}} U^{2} \delta_{t}(0) + O\left(\left[ \frac{\Nx}{\nqw\Nel}\right]^4 \right).\quad\quad
	\label{eq:rate}
\end{eqnarray}

With the same notations, the contribution of the many-body physics to the lifetime of the initial state is\cite{Combescot07}
\begin{eqnarray}
	\frac{t}{T} & = & \braket{\tilde{\psi}_{t}}{\tilde{\psi}_{t}} - \left| \langle \psi_{i} \big| \tilde{\psi}_{t} \rangle \right|^{2} .
\end{eqnarray}
The second term can be calculated using the same method as for the transition rate and is found to contribute only to higher orders in the perturbation.
After some rather cumbersome algebra, the first term is found to be of the form
\begin{eqnarray}
	\frac{1}{T} & = & \frac{1}{2} \sumk{p} \Gamma_{\mathbf{p}} ,
	\label{eq:lifetime}
\end{eqnarray}
where it is implicitly assumed that the summation is restricted to small wavevectors.
To obtain this result we made $F_{t}$ act at zero-th order and we used the fact that bright and dark intersubband excitations are not resonant. The counterintuitive $1/2$ factor comes from  the overcompleteness of the  composite boson basis.

%%%%%%
\subsection{Bosonic case}

In this paragraph we calculate the same quantities as in the previous one but we assume that intersubband excitations are bosons and that their dynamics is described by Hamiltonian $H_{B}$. We can therefore use a traditional Fermi golden rule.

The initial and final states in this case are 
\begin{eqnarray}
\ket{\psi_{i}}&\propto& B_{\mathbf{q}}^{\dagger \, \Nx} \ket{G}
\end{eqnarray}
and
\begin{eqnarray}
\ket{\psi_{f}}&\propto& \Bc{q+p} \Bc{q-p} B_{\mathbf{q}}^{\dagger \, \Nx-2} \ket{G}.  \nonumber
\end{eqnarray}
The transition rate is thus given by the formula
\begin{eqnarray}
	\Gamma_{\mathbf{p}} & = & \frac{2\pi}{\hbar} \frac{\Nx (\Nx-1)}{\nqw^{2} \Nel^{2}} U^{2} \delta_{t}(0) ,
\end{eqnarray}
which is the same as in \Eq{eq:rate} up to third order in $\Nx / (\nqw\Nel)$.
As all bright intersubband excitations are resonant the lifetime of the initial state is the sum of the transition rates
\begin{eqnarray}
	\frac{1}{T} & = & \sumk{p} \Gamma_{\mathbf{p}} ,
\end{eqnarray}
where the summation is again restricted to small wavevectors. A comparison with \Eq{eq:lifetime} shows that this method underestimates the true lifetime by a factor two. 
This is coherent with the results in Ref. [\onlinecite{Combescot07}], that show how an effective Hamiltonian approach that correctly calculates transition rates needs to take into account an {\it ad hoc} factor $1/2$ when calculating lifetimes, due to the overcompleteness of the composite boson basis. This is of course simply implies a renormalization of the composite boson density of states.

\end{document}